\documentclass [12pt]{article}
\usepackage {amssymb}
\usepackage {amsmath}
\usepackage {graphicx}
\usepackage {longtable}

\begin{document}

\begin{flushleft}\footnotesize{ {\it Kybernetes: The Intern.
Journ. of Systems \& Cybernetics} {\bf 32}, No. 7/8, pp. 976-1004
(2003) \\
 {\rm (a special issue on new theories about time and space,
Eds.: L. Feng, B. P. Gibson and Yi Lin)}}
\end{flushleft}

\section*{Scanning the structure of ill-known spaces: Part 2. \\
Principles of construction of physical space}

\medskip

\begin{center}
\textbf{Michel Bounias$^{(1)}$ and Volodymyr
Krasnoholovets$^{(2)}$}
\end{center}

\begin{center} ($^{1}$) BioMathematics Unit(University/INRA, France, and
IHS, New York, USA), \\ Domain of Sagne-Soulier, 07470 Le Lac
d'Issarles, France            \\
($^{2}$) Institute of Physics, National Academy of Sciences, \\
Prospect Nauky 46,     UA-03028, Ky\"{\i}v, Ukraine
\end{center}

\textbf{Abstract.} An abstract lattice of empty set cells is shown
to be able to account for a primary substrate in a physical space.
Space-time is represented by ordered sequences of topologically
closed Poincar\'{e} sections of this primary space. These mappings
are constrained to provide homeomorphic structures serving as
frames of reference in order to account for the successive
positions of any objects present in the system. Mappings from one
to the next section involve morphisms of the general structures,
standing for a continuous reference frame, and morphisms of
objects present in the various parts of this structure. The
combination of these morphisms provides space-time with the
features of a nonlinear generalized convolution. Discrete
properties of the lattice allow the prediction of scales at which
microscopic to cosmic structures should occur. Deformations of
primary cells by exchange of empty set cells allow a cell to be
mapped into an image cell in the next section as far as mapped
cells remain homeomorphic. However, if a deformation involves a
fractal transformation to objects, there occurs a change in the
dimension of the cell and the homeomorphism is not conserved. Then
the fractal kernel stands for a "particle" and the reduction of
its volume (together with an increase of its area up to infinity)
is compensated by morphic changes of a finite number of
surrounding cells. Quanta of distances and quanta of fractality
are demonstrated. The interaction of a moving particle-like
deformation with the surrounding lattice involves a fractal
decomposition process that supports the existence and properties
of previously postulated inerton clouds as associated to
particles. Experimental evidence of the existence of inertons are
reviewed and further possibilities of experimental proofs are
proposed.

\bigskip
\textbf{Key words:} origin of particles; origin of quantum
property; origin of gravitation; cosmic features from microscale
properties; nothingness singleton.

\bigskip
\textbf{PACS classification:} 02.10.Cz; -- set theory; 02.40.Pc --
general topology; 03.65.Bz -- foundations, theory of measurement,
miscellaneous theories.

\newpage

\section*{1. Introduction}

\hspace*{\parindent} Former parts of this study were dealing with
some founding principles about how to assess in the more accurate
though general way possible how one can define the space of magmas
(that is, the sets, combination rules and structures) in which a
given proposition can be demonstrably shown to be valid (Bounias,
2001; Bounias and Krasnoholovets, 2003). Such a space, when
identified, is called a probationary space (Bounias, 1997; 2001).
Here, it will be presented the formalism which leads from
existence of abstract (e.g. purely mathematical) spaces to the
justification of a distinction between parts of a physical space
that can be said empty and parts which can be considered as filled
with particles. This question is thus dealing with a possible
origin of matter and its distribution, and changes in this
distribution gives raise with motion, that is with physics, in the
sense pointed by de Broglie and by Dirac (see Rothwarf, 1998).
Experimental evidence and propositions for further verifications
will then be presented and discussed.

Recent findings (Krasnoholovets and Ivanovsky, 1993;
Krasnoholovets 1997) in the realm of fundamental physics supports
the prediction that an abstract lattice whose existence originates
in the existence of the empty set, is able to correctly account
for various properties of our observed space-time at both
microscopic and cosmic scales. The model of Krasnoholovets and
Byckov (2000), Krasnoholovets (2000), proposed a new research
methodology based on some practical standpoints. Specifically (see
e.g. Okun, 1988), the values of the constants of electromagnetic,
weak and strong interactions as functions of distances between
interacting particles converge to the same at a scale about
10$^{-28}$ cm. This suggests that a violation of space homogeneity
took place at this size. The model proceeds from the assumption
that all quantum theories (quantum mechanics, electrodynamics,
chromodynamics and others) are in fact only phenomenological:
accordingly, for the understanding of real processes occurring in
the real microworld, one needs a submicroscopic approach which, in
turn, should be available for all peculiarities of the
microstructures of real space. In other terms, gauges for the
analysis of all components of the observable universe should
belong to an ultrafilter, as argued in Part 1 of this study. The
investigations about the model of inertons (Krasnoholovets, 1997,
2000; 2001a,b,c) has suggested that a founding cellular structure
of space shares discrete and continuous properties, which is also
shown to be consistent with the abstract theory of foundations of
existence of a physical space (Bounias and Bonaly, 1997).

\section*{2. Preliminaries}

\subsection*{2.1. About gaps in former assessment of probationary
spaces}

\noindent \textbf{2.1.1. Quantum mechanics.} Quantum mechanicsis
founded on the calculation of the probability that a particle is
present in a given volume of space. This theoretical approach
postulates the existence of undefined objects called corpuscles,
and does not state about the structure and properties of any
embedding medium, which is even considered as forbidden
(Blokhintsev, 1981). Only recently, however, was raised the need
that this medium, sometimes called the void, should be a space
allowing the formation of pairs of particles and antiparticles
(see Boyer, 2000, for review), so as to justify for the existence
of a material world. However, this postulate just displace the
question of the corresponding embedding medium, which is supposed
not to exist independently from the photons but is often
considered as if it was independent from at least large matter
masses. Parameter time is not basically but implicitly present in
the foundations of quantum physics. The concept of velocity of
wave propagation and its expressions in the uncertainty principle,
the Bell's inequalities, etc. emerge from further developments of
the theory.

\medskip
\noindent \textbf{ 2.1.2. Relativity.} The relativistic theory
postulates the existence of frames of references and of the
validity of some particular cases of measure, used as classical
metrics, still without consideration for the embedding medium
(Blokhintsev, 1976). It also postulates the primary existence of
parameter time and the consistency of the possibility of motion in
an undefined space, sometimes identified with "void", with the
properties of this "void". However, the limitation found for the
velocity of light up to the cosmological constant through
electromagnetism, and the proposition of curvature of space
implicitly impose some conditions on some relevant embedding
medium (Einstein, 1920; Marinov, 1996; Keilman, 1998, and many
others).

\subsection*{2.2. Assessment of existence of a space-time-like
structure}

\hspace*{\parindent} A former conjecture (Bonaly, 1992) stated
that a characteristic of a physical space is that it should be in
some way observable. This implies that an object called the
"observer" should be able to interact with other objects said
"observed". In order to make no confusion with the usual
vocabulary of systems theory, we will instead refer to the
"perceiver" and the "perceived" objects. The conjecture implied
that perceived objects should be topologically closed, otherwise
they would offer no frontier to allow a probe to reflect their
shape. Therefore, the first step of the work was to assess the
existence of closed topological structures, and a proof was given
that the intersection of two spaces having nonequal dimensions
owns its accumulation points and is therefore closed. We propose
here a shorter alternative proof.

\medskip
\noindent  \textbf{Theorem 2.2.} The intersection of two connected
spaces with nonequal dimension is topologically closed.

\medskip\noindent
\underline{Alternative proof.} Let $\rm E^{{\kern 0.3pt} n}$ and
$\rm E^{{\kern 0.3pt}m}$ be two spaces with topological dimensions
n, m ($\rm m>n$) embedded in $\rm W^\infty$, a compact connected
space. Call $\rm S^{{\kern 0.3pt}n}$ the intersection $\rm
E^{{\kern 0.3pt} n} \cap E^{{\kern 1pt} m}$ and call $\rm X^{n}$
the complementary of S in $\rm E^{{\kern 0.5pt} n}$. Consider the
continuity of mappings in $\rm W^\infty$ inducing continuity to
$\rm E^{{\kern 1pt}n}$ and $\rm E^{{\kern 0.3pt}m}$: the
neighborhood of any point in $\rm E^{{\kern 0.3pt}m}$ is the
mapping of a neighborhood of a point in $\rm X^{n}$. Suppose $\rm
S^{{\kern 0.3pt}n}$ is open: then, since the union of open sets is
open, then the entire $\rm E^{{\kern 0.3pt} m}$ is neighborhood of
any point in $\rm S^{{\kern 0.7pt}n}$. Thus, there would exist a
bijective mapping of opens of $\rm X^{{\kern 1pt} m}$ on opens of
$\rm S^{{\kern 0.7pt} n}$. In particular, a open subset of $\rm
(n+2)$ points in $\rm S^{{\kern 0.7pt} n}$ could be homeomorphic
to a $\rm (n+l)$-simplex in $\rm E ^{{\kern 0.3pt} m}$. This is
impossible because two spaces with nonequal dimensions cannot be
homeomorphic. Therefore, $\rm S^{{\kern 1pt} n}$ is closed.

The closed 3-D intersections of parts of a n-space (with $\rm n
\leq 3$) own the properties of Poincar\'{e} sections (Bonaly and
Bounias, 1995). Then, given a manifold of such sections, the
mappings of one into another section provide an ordered sequence
of corresponding spaces in which closed topological structures are
to be found: this accounts for a time-like arrow. Since the
Jordan-Veblen theorem states that any path connecting the interior
of a closed to an outside point has a nonempty intersection with
the frontier of the closed, interactions between closed objects
are allowed: this accounts for physical interactions. Furthermore,
if such a path is connected to a converging sequence of mappings,
the fixed points (of Banach type, here) will stand for perceptions
of the outside. Moreover, since the Brouwer's theorem states that
in a closed, all continuous mappings have a fixed point, and that
the brain represents a compact complete space in which mappings
from a topological into a discrete space are continuous, there
exists an associate set of fixed points (of the Brouwer's type)
standing for the self (Bounias and Bonaly, 1997). Finally, spaces
of topologically closed parts account for interaction and for
perception, thus they meet the properties of physical spaces.

In a former conjecture, Bonaly and Bounias (1993) proposed that
the fundamental metrics of our space-time should be represented by
a convolution product where the embedding part U4 would be
described by the following relation:
 $$
{\rm U}4 = \int \Big( \int _{\rm dS}({\rm d} {\vec {\rm x}} \cdot
{\rm d} {\vec {\rm  y}} \cdot {\rm d} {\vec {\rm z}})\ast {\rm d}
\Psi ({\rm w}) \Big) \eqno(2.1)
 $$

\noindent where dS is an element of space-time, and $\rm d\Psi
(w)$ a function accounting for the extension of 3-D coordinates to
the 4th dimension through convolution ($\ast$) with the volume of
space. Formal proofs of this structure will be provided below.

\section*{3. On foundations of space-time}

\subsection*{
2.1. Space-time as a topologically discrete structure}

\hspace*{\parindent} How two Poincar\'{e} sections are mapped is
assessed by using a natural metrics of topological spaces: the
set-distance, first established for two sets (Bounias and Bonaly,
1996) and further generalized to manifolds of sets (Bounias,
1997), Figure 1. In brief: let $\rm \Delta (A,B,C,...)$ the
generalized set distance as the extended symmetric difference of a
family of closed spaces:
 $$
\rm \Delta(A_{i})_{{\kern 1pt}i {\kern 1pt} \in {\kern 1pt}
N}={\mathop \complement_{\cup \{ Ai \}}}{\kern 1pt} {\mathop \cup
_{{\kern 2pt} i\neq j}}{\kern 1pt}(A_i \cap A_j).
 \eqno(3.1)
 $$

\begin{figure}
\begin{center}
\includegraphics[scale=1.6]{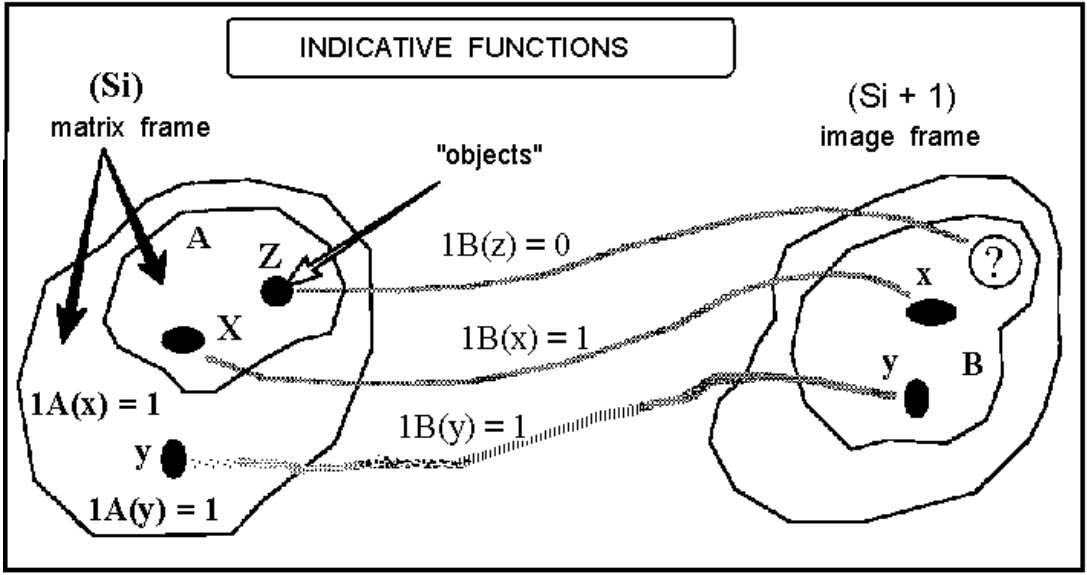}
\caption{\small{The mappings of a Poincar\'{e} section S(i) into a
section S(i+1) imposes the conservation of the topologies of the
general structure of the mapped spaces, so that changes in the
position of objects situated inside these structures can be
characterized. A closed should be mapped into an equivalent
closed, an open into an equivalent open. The situation of points
(x, y, ...) with respect to these reference structures (A, B, ...)
are described by indicatrix functions 1A(x), 1B(x), ... }}
\label{Figure 1}
\end{center}
\end{figure}

The complementary of $\Delta$, that is $\rm \cup _{{\kern 2pt}
i\neq j}(A_i \cap A_j)$ in a closed space is closed. It is also
closed even if it involves open components with nonequal
dimensions. Thus, in this system, $\rm \mathfrak{m} \langle \{
A_i\} \rangle = \cup _{{\kern 2pt} i\neq j}(A_i \cap A_j)$ has
been called the "instant", that is the state of objects in a
timeless Poincar\'{e} section (Bounias, 1997). Since distances
$\Delta$ are the complementaries of objects, the system stands as
a manifold of open and closed subparts. Mappings of these
manifolds from one into another section which preserve the
topology stand for a reference frame in which the "analysis situ"
(the original name for "topology") will allow to characterize the
eventual changes in the configuration of some components: if
morphisms are observed, then this will be interpretable as a
motion-like phenomenon when comparing the state of a section to
the state of the mapped section.

It should be noted that the spaces referred above can exist upon
acceptance of the existence of the empty set as a primary axiom
(Bounias and Bonaly, 1997), with consequences which will be
addressed below.

\medskip
\noindent \textbf{Lemma 3.1.} The set-distance provides a set with
the finer topology and the set-distance of nonidentical parts
provides a set with an ultrafilter.

\medskip\noindent \underline{Proof.} The set-distance is founded on
$\{ \cap{\kern 1pt} \cup \in \}$ and it suffices to define a
topology since union and intersection of set-distances are
distances, including $\rm \Delta (A,A) = \O$. The latter case must
be excluded from a filter, which is nonempty. Then, since any
filter and any topology is founded on $\{ \cap {\kern 3pt}$$\cup$
$\in$${\kern 1pt}\notin$$\supset\}$, it is provided with $\Delta$.
Conversely, regarding a topology or a filter founded on any
additional property ($\perp$), this property is not necessarily
provided to a $\Delta$-filter. The topology and filter induced by
$\Delta$ are thus respectively the finer topology and an
ultrafilter.

\begin{figure}
\begin{center}
\includegraphics[scale=2.3]{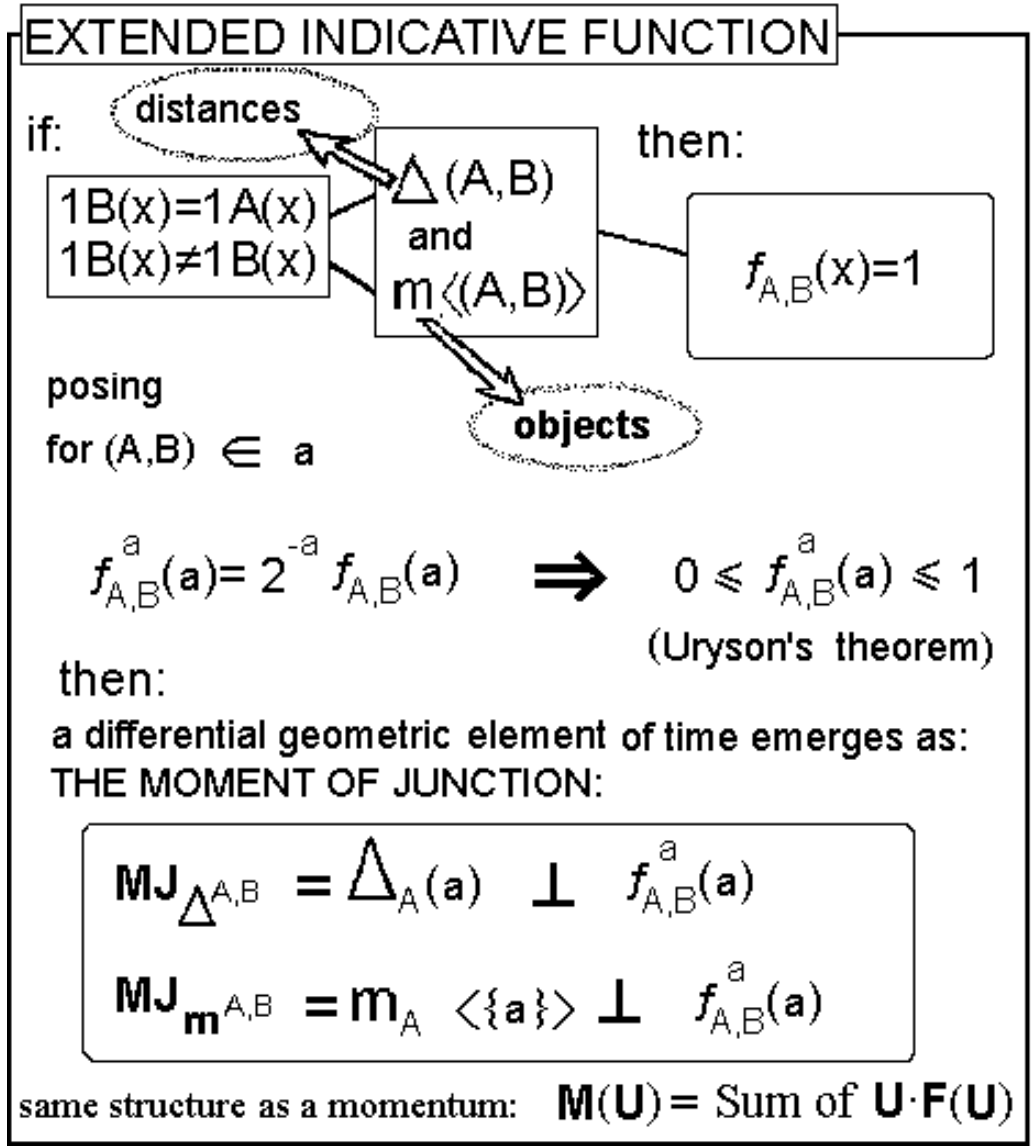}
\caption{\small{The composition of indicative functions of the
position of points within the topological structures used as
frames of references provides an extended indicative function. The
composition of pairs of indicative functions provides a new
function ($f$) indicating the changes occurring in the situation
of objects. The composition of the topological distances $\rm
\Delta (A,{\kern 1pt}B, {\kern 1pt}...) = \complement_{A\cup B ...
}(A\cap B \cap ...)$ or the topological "instances" ($\rm
\mathfrak{m} \langle (A,{\kern 1pt}B,{\kern 1pt}C, {\kern
1pt}...)\rangle = (A\cap B)\cup (A \cap C)\cup (B\cap C)$ with
function $f$ over the populations of objects in the considered
sets leads to a momentum-like structure called "Moment of
Junction" and accounting for elements of the differential geometry
of space.}} \label{Figure 2}
\end{center}
\end{figure}

The mappings of both distances and instances from one to another
section can be described by a function called the "moment of
junction", since it has the global structure of a momentum (Figure
2). Consider the particular case of the homeomorphic sequence of
mappings of the general topology of the system: this provides a
kind of reference frame, in which it will become possible to
assess the changes in the situation of points and sets of points
eventually present within these structures. The origin of such
points will be addressed in the next section. Here, just consider
point (x, y, z, ...) belonging to either of closed and open parts.
For any x belonging to a set Ei in a section S(i), an indicatrix
function 1(x) is defined by the correspondence of x with some c(x)
in S(i+1):
 $$
\rm x \in E(i), 1_{{E{\kern 0.3pt}i}}(x)=1 \ iff \ x \in E{\kern
0.3pt}i; \ 1_{{E{\kern 0.3pt}i}}(x) =0 \ iff \ x \notin Ei;
 $$
 $$
\rm c(x) \in E(i+1), \ 1_{{(E{\kern 0.3pt}i+ 1)}}(x)=l \ iff  \
c(x) \in E(i+l); \quad \
 $$
 $$
\rm 1_{{(E{\kern 0.3pt}i+ 1)}}(x) = 0 \ iff \ c(x) \notin E(i+1).
\qquad\qquad\qquad\qquad \
 \eqno(3.2)
 $$

Then, a function $\rm {\it f}_{{(E{\kern 0.3pt}i,{\kern 1.5pt}
E{\kern 0.3pt}i + 1) } }$ (more shortly noted $\rm {\it f}_{
{\kern 1pt}E}(E)$) is defined as:
 $$
\rm {\it f}_{{\kern 0.7pt}\rm E} =1 \ iff: 1_{{\kern
0.5pt}(E{\kern 0.3pt}i)} = 1_{(E{\kern 0.3pt}i+1)},
 $$
 $$
\rm {\it f}_{{\kern 0.7pt}\rm E} = 0 \ iff: 1_{(E{\kern 0.3pt}i})
\neq 1_{(E{\kern 0.3pt}i+1)}.
 \eqno(3.3)
 $$

Summing over all points x in the whole of \{E\} provides $\rm {\it
f}_{{\kern 1pt} E}(E)$ that accounts for a distribution of the
indicatrix functions of all points out of the maximum number of
possibilities, which would be $\rm 2^{{\kern 1pt}E}$ for the set
of parts of set E. This finally leads to the expression of the
proportion of points involved in the mappings of parts of $\rm
E(i+1)$ into
 $$
\rm {\it f} _{{\kern 0.5pt}E}^{{\kern 0.5pt}E} ={\it f}_{{\kern
0.5pt}E}^E(E)/2^{{\kern 1pt}E}, \ \ \ 0<{\it f}_E^E(E)<1.
 \eqno(3.4)
 $$

  Two species of the moment of junction are represented by the
composition ($\perp$) of $\rm {\it f} _{{\kern 0.5pt}E}^E$ with
either the set-distance of the instance, since $\rm E\supseteq
\Delta  (E) \cup \mathfrak{m} \langle E\rangle$ and the
distribution of points in the complementary structures is not the
complementary of their distributions. Hence:
 $$
\rm MJ_{\Delta}= \Delta(E)\perp {\it f}^{{\kern 0.5pt}E}_{{\kern
0.5pt}E}(E),
 \eqno(3.5{\kern 0.4pt}\rm a)
 $$
 $$
\rm MJ_{\mathfrak{m}} = \mathfrak{m} \langle {\kern 1pt} E {\kern
1pt} \rangle \perp {\it f}^{{\kern 0.5pt}E}_{{\kern 0.5pt}E}(E).
 \eqno(3.5{\kern 0.4pt}\rm b)
 $$

Generally, one will have $\rm MJ_{\Delta} \neq MJ_{\mathfrak{m}}$.

As a composition of variables with their distribution, relations
(3.5{\kern 1pt}a,b) actually represent a form of momentum.

\subsection*{3.2. Space-time as fulfilling a nonlinear convolution
relation}

\hspace*{\parindent} The "moments of junctions" (MJ) mapping an
instance (a 3-D section of the embedding 4-space) to the next one
apply to both the open (the distances) and their complementaries
the closed (the reference objects) in the embedding spaces. But
points standing for physical objects able to move in a physical
space may be contained in both of there reference structures.
Then, it appears that two kinds of mappings are composed with one
another.

\medskip
\noindent \textbf{Theorem 3.2.} A space-time-like sequence of
Poincar\'{e} sections is a nonlinear convolution of morphisms.

\medskip \noindent
\underline{Proof.} The demonstration involves the following four
steps. \\

\noindent (i) One kind of mapping $\mathcal{M}$ connects a frame
of reference to the next one: here, the same organization of the
reference frame-spaces must be found in two consecutive instants
of our space-time, otherwise, no change in the position of the
contained objects could be correctly characterized.

However, there may be some deformations of the sequence of
reference frames, on the condition that the general topology is
conserved, and that each frame is homeoporphic to the previous
one. Mappings $\mathcal{M}$ will thus denote the corresponding
category of morphisms.  \\

\noindent (ii) The other kind of mapping ($\mathcal{J}$) connects
the objects of one reference cell to the corresponding next one.
Mappings ($\mathcal{J}$) thus behave as indicatrix functions of
the situation of objects within the frames, and therefore, they
are typically relevant from the "Analysis situs", that is the
former name for topology, originally used by Poincar\'{e} himself
(Bottazzini, 2000).

These morphisms thus belong to a complementary category.

Then, each section, or timeless instant (that is, a form of the
above more general "instance") of our space-time, is described by
a composition ($\bigcirc$) of these two kinds of morphisms:
 $$
{\rm Space-time} \rm \  instant = (\mathcal{M} \bigcirc
\mathcal{J}).
 \eqno(3.6)
 $$  \\

\noindent (iii) Stepping from one to the next instant is finally
represented by a mapping $\mathbf{T}$, such that the composition
$(\mathcal{M} \bigcirc \mathcal{J})$ at iterate (k) is mapped into
a composition $(\mathcal{M} \perp \mathcal{J})$  at iterate (k+i):
 $$
\rm (\mathcal{M} \perp \mathcal{J})_{{\kern 1pt}k+i} =
\mathbf{T}^{{\kern 0.3pt}\perp} (\mathcal{M} \bigcirc
\mathcal{J})_{{\kern 1pt}k}.
 \eqno(3.7)
 $$

Hence, mapping ($\mathbf{T}^{{\kern 0.3pt}\perp}$) appears like a
relation of the type $\rm \mathcal{R}_{{\kern 1pt} k+i}$ similar
to that denoted below by $\rm \mathcal{R}_{{\kern 1pt} k+j}$ that
maps a function $\rm F_{i+k}$ into  $\rm F_{j+k}$:
 $$
\rm F_{j+k}^\prime =\mathcal{R}_{{\kern 1pt}(k+j)}F_{{\kern
1pt}i+k}. \eqno(3.8)
 $$  \\

\noindent (iv) The above relation represents a case of the
generalized convolution, that is a nonlinear and multidimensional
form of the convolution product, which has been first described by
Bolivar-Toledo et al. (1985). The authors have proposed this
concept as a tool for computing the behavior of visual perception.
The demonstration that relation (3.8) is a form of a convolution
is achievable by taking the following example.

Let $\rm \alpha_{{\kern 1pt}(j-k)}$ a particular form of $\rm
\mathcal {R}_{{\kern 1pt}(k+j)}$; then equation (3.8) becomes:
 $$
\rm F^\prime_{{\kern 1pt} k} = \sum \alpha_{{\kern 1pt} (j-k)}
F_j,
 \eqno(3.9)
 $$
that is, for the case of an integrable space
 $$
\rm F^\prime (X) = \int \alpha (X^\prime - X) F(X^\prime)
d(X^\prime).
 \eqno(3.10)
 $$
So the relation exhibits a great similarity with a distribution of
functions, in the Schwartz sense (Schwartz, 1966):
 $$
\langle f, {\kern 1pt} \varphi \rangle = \rm \sum \varphi (x) {\it
f} (x) d x
 \eqno(3.11)
 $$
or a convolution product
 $$
\rm \int_{E} {\it f}(X - u){\kern 1pt} F(u){\kern 1pt}  d(u) =
({\it f} \ast F){\kern 1pt}(X).
 \eqno(3.12)
 $$

Thus, the connection from the abstract universe of mathematical
spaces and the physical universe of our observable space-time is
provided by a convolution of morphisms, which supports the
conjecture of relation (2.1).

Interestingly, the distributions were primarily considered by
Schwartz as an invention, in contrast with other concepts which he
considered as the discovery of preexisting foundations of the
total universe (Schwartz, 1997). Now, the present work could
provide the distributions with the status of a discovery.

\medskip\noindent
\textbf{Remarks 3.2.} Our observable space-time should possess
nonlinear properties which necessarily involve specific features
(Lin and Wu, 1998): therefore, peculiar approaches are needed
through appropriate mathematical concepts, otherwise incorrect
descriptions of the real world result as a consequence of
approximate treatment of nonlinear model resolution (Wu and Lin,
2002).

\section*{4. Relative scales in the empty-set lattice}

\hspace*{\parindent} It has been demonstrated in Bounias and
Krasnoholovets (2003) that the antifounded properties of the empty
set provide existence to a lattice involving a tessellation of the
corresponding abstract space with empty balls. This structure has
thus been called a "tessellattice". Its formalism will be
completed in section 3.2 up to Corollary 3.2.5.

\subsection*{4.1. Quantum levels at relative scales}

\hspace*{\parindent} Inside any of the above spaces, properties at
micro-scale are provided by properties of the spaces whose members
are empty set units. It will be shown here that particular levels
of a measure of these units can be discerned.

\medskip
\noindent\textbf{Lemma 4.1.2.} The Cartesian product of a finite
beginning section of the integer numbers provides a variety of
nonequal empty intervals.

\medskip
\noindent \underline{Proof.} Let $\rm A_{(N+1)} = \{0,{\kern 2pt}
1, {\kern 2pt}2, {\kern 2pt}...,{\kern 2pt} N\}$ denote a
beginning section, that is the set of all members of a part
$(M,\preceq)$ of the natural integers ({\bf N}) provided with an
order relation ($\preceq$), which are lower than (N+1). A set
\{\O, a,b,...\} is equipotent with \{0, 1,2,...,n\} and will be
denoted by ($\rm E _{{\kern 1pt}n}$). Then, since any set contains
\O, one has ($\rm E_{{\kern 1pt}n}) \equiv A_{(N+1)}$.

Consider now $\rm (E_{{\kern 1pt}n})^{2} = (E_{{\kern 1pt}n})
\times (E_{{\kern 1pt}n})$. The resulting set contains ordered
pairs including the diagonal (aa, bb, ...) and the heterologous
pairs (ab, ac, bd, ...) accounting for rational numbers ($\rm aa
\mapsto 1$, $\rm ab \mapsto a/b$, $\rm dc \mapsto d/c$, etc.).
Since all members of the diagonal are mapped into 1, there remains
$\rm n^{2}-(n-l)$ distinct pairs. A rational can be represented by
a 2-simplex or facet, whose small sides are the corresponding
integers. Jumping to three dimensional conditions with $\rm
(E_{{\kern 1pt}n})^{3} = (E_{{\kern 1pt}n})^2 \times (E_{{\kern
1pt}n})$, each new rational is represented by a 3-simplex. This
representation offers the advantage over the usual square to cube
representation of avoiding several squares or cubes to share
common edges or facets. Then, the number of 3-simplexes reflects
the number $\rm \mathcal{R}(E_{{\kern 1pt}n})^{3}$ of rational
numbers available from any initial beginning segment $\rm
A_{{\kern 1pt}{N+1}}$.

Consider now the cuttings or segments, represented by intervals
between any two of these rational numbers. These segments
represent the whole of available distances in the corresponding
subpart of a 3-space. Remind that such a subpart is involved in
some part of a ordered sequence, that is in a segment of an
observable space-time.

Let [0, n] be the larger interval of ($\rm E_{{\kern 1pt}n}$).
Denote by $\rm \mathfrak{m}_E]0, {\kern 2pt}n[$ a measure of the
open part or interior ]0, n[ of this interval in space ($\rm
E_{{\kern 1pt}n}$): if E is a segment of N, one has:
 $$
\rm \mathfrak{m}_E]0,{\kern 2pt} n[ = {\O}^{N}.
 \eqno(4.1)
 $$

This is also encountered in the Cartesian product ($\rm E_{{\kern
1pt}n})^{3}$ when it includes (\O).

Consider now any of the smaller intervals ($\sigma$) in $(\rm
E_{{\kern 1pt}n})^{3}$ and denote by $\rm
\mathfrak{m}_E[(\sigma(E_{{\kern 1pt}n})^{3}]$ its measure. By
definition:
 $$
\rm \mathfrak{m}_E[(\sigma(E_{{\kern 1pt} n})^{3}] <
\mathfrak{m}_E[0,{\kern 2pt} n].
 \eqno(4.2)
 $$

Call \O$^{\rm Q}$ any of the open or interior $\rm
](\sigma(E_{{\kern 1pt}n})^{3}[$, then since $\rm G\subset
H\Rightarrow (g\subset G) \subset (H)$ and that the distance of
the interior of a set to its frontier is naught (Tricot, 1999):
 $$
\rm {\O}^Q \subset {\O}^N.
 \eqno(4.3)
 $$

This imposes an order relation holding on empty sets constructed
on various segments of a finite product of finite beginning
sections of the set of natural integers or equipotent to such a
section.   \qquad\qquad\qquad \qquad\qquad\qquad\qquad\qquad
(Q.E.D.)

\medskip\noindent
\textbf{Corollary 4.1.2.} A finite set of rational numbers
inferring from a Cartesian product of a finite beginning section
of integer numbers establishes a discrete scale of relative sizes.

\medskip\noindent
\underline{Proof.} (i) Intervals are constructed from mappings
$\mathcal{G}$: $\rm \mathbf{N}^{D}\mapsto \mathbf{Q}$ of
($\mathbf{N} \times \mathbf{N} \times \mathbf{N} \times \mathbf{N}
\times ...$) in $\mathbf{Q}$. For example, with $\rm D=2$, the
smaller ratios available are 1/n and $\rm 1/(n-l)$, so that their
distance is the smaller interval: $\rm 1/\big(n{\kern
0.5pt}(n-1)\big)$. Consider now the few smaller intervals
($\sigma$) in $(\rm E_{{\kern 1pt}n})^{3}$. One can observe, in
the following order of increasing sizes:
 $$
\rm \forall{\kern 1pt} n>1:
\qquad\qquad\qquad\qquad\qquad\qquad\qquad\qquad\qquad
\qquad\qquad\qquad\qquad\qquad\qquad
 $$
 $$
\rm (\sigma_{({\kern 0.3pt}i)})=\frac 1{n^2(n-1)} <
(\sigma_{({\kern 0.3pt}ii)}) = \frac 1{n(n-1)^2} <
(\sigma_{({\kern 0.3pt}iii)}) = \frac {(n-1)}{n^2(n-l)^2}.
 \eqno(4.4)
 $$

Let us consider maximal of the ratios of larger (n) to smaller
$\rm \big(1/n^{2}(n-l)\big)$ segments; then one gets
 $$
\rm \frac{\max(\sigma)} { min(\sigma)} = n^{3}(n-l).
 \eqno(4.5)
 $$

One possibility of a scaling progression covering integer
subdivisions (n) consists in dividing a fundamental segment ($\rm
n=l$) by 2, then each subsegment by 3, etc. Thus the size of
structures is a function of iterations (n). At each step
($\nu_{\rm j}$) the ratio of the size in the dimension D will be
$(\Pi \nu_{\rm j})^{\rm D}$, so that the maximal ratio will be,
following (4.5),
 $$
\rm \varrho \propto \Big\{ (\Pi \nu_{j})^D {\kern 1pt} \big( \Pi
\nu_{j}-1 \big) \Big\}_{{\kern 1pt}j{\kern 1pt}=1 {\kern
1pt}\rightarrow {\kern 1pt} n}. \eqno(4.6)
 $$

The manifold $(\Pi \nu_{\rm j})$ is a commutative
Bourbaki-multipliable indexed on the integer section $\rm
I=[1,{\kern 1pt}n]$. In practice, values can be written as $\rm
\varrho _j = a_j. {\kern 1pt}10^{x_j}$ where in the base 10 one
will take $\rm a_j.$ belonging to the (always existing)
neighborhood of unity, i.e. $\rm a_j. \in ]1[$, and look at the
corresponding integer exponents $\rm x_j$ as the order of sizes of
structures constructed from the lattice $\rm \mathcal{L}=(\Pi
\nu_j)^{D}$. Regarding distances ($\rm D=1$) to areas and volumes
($\rm D=2$ and 3) equation (4.6) consistently provides the
predicted orders of scales listed in Table 1. The latter represent
quantic-like levels of clusters of objects sharing successive
orders of sizes while constructed one from the others.

\medskip\noindent
(ii) There is a finite number of segments or intervals, since
their supremum is the number of 3-simplexes, which is finite, and
contains redundant terms. This number is the number of pairwise
combinations of distinct rationals. Therefore:
 $$
\rm sup {\kern 1pt} \mathcal{N}{\kern 1pt}(E_{{\kern 1pt} n})^{3}
= \mathbf{C}^{{\kern 1pt}2}_{n^2-n+1} =\tfrac 12 {\kern 1pt}{\kern
1pt} n(n-l)(n^2-n+1).
 \eqno(4.7)
 $$
A general formula for exactly ($\rm \mathcal{N} (E_{{\kern 1pt}
n})^{3}$) is not readily available since it involves prime numbers
occurring in ($\rm E_{{\kern 1pt} n}$), but this does not change
the meaning of the reasoning.

Finally, there is a finite number of ratios of segment sizes
imposing on a subpart \newline of a space-time sequence a limited
number of relative scales for any of the objects \newline
represented by closed subspaces in $\rm (E_n)^{3}$.
\qquad\qquad\qquad \qquad\qquad (Q.E.D.)

\medskip \noindent
\textbf{Corollary.} The axiom of availability (stating that a rule
provided to a set must be considered as explicitly applying to the
set's members and parts) is necessary to an exploration of an
unknown space, either mathematical or physical.

The simple following counter-example provides the proof. Suppose
the axiom of availability is not stated: then, a complete subset
of the rational numbers may not be provided in all bases by the
Cartesian product of a segment of natural numbers.

Let $\rm E_{{\kern 0.3pt} n} = \{1,{\kern 1pt}2, {\kern 1pt}
...,{\kern 1pt} n\}$ with $\rm n<9$, and let two integers, $\rm
p,{\kern 2pt}q \in E_{{\kern 0.3pt}n}$. Then, the pair $\rm
(p,{\kern 1pt}q)\in (E_{{\kern 0.3pt}n}^{{\kern 0.3pt}2}) =
E_{{\kern 0.3pt}n} \times E_{\kern 0.3pt n}$. Usually, $\rm
(p,{\kern 1pt}q)$ accounts for the ratio p/q, so that the set of
pairs $\rm (p,{\kern 1pt}q)$ is equipotent to the set of rational
numbers noted as fractions: ($\rm e_i.{\kern 1pt}d_1d_2...d_i$)
where $\rm e_i$. stands for the entire part and $\rm
d_1d_2d...d_i{\kern 1pt}...$ for the decimal part.

Let $\rm n=4$, and take $\rm p=1$, $\rm q=4$. Then the ordered
pair (1,4) stands for the ratio 1/4. However, writing 1/4 = 0.25
needs digit 5 to be available, whereas one has just $1,2,3,4$
available, not 5. Therefore, in this system, since digit 5 does
not exist unless the additional axiom of the addition is
introduced, the mapping of ordered pairs to the writing in base 10
of the corresponding rational numbers is not valid. The
availability of the power set of parts, i.e. the infinitely
iterated sets of parts of the sets of parts (Bounias, 2001) is
enough to break this barrier.

\bigskip

\textbf{Table 1.} Range and intermediate levels of the scale of
size of objects composing a universe constructed as described in
relation (4.6). Intervals constructed with powers of 10 on the
neighborhood of unity $\rm (a ]1[{\kern 1pt})$ are confronted
through dimensions: (i) $\rm D=1$ with the simple multipliable set
$\rm {\Pi}_j (\nu_j)$; (ii) $\rm D=2$ involving intervals $\rm
l/n(n-l)$; (iii) $\rm D=3$ involving intervals $\rm 1/n^{2}(n-l)$.
Here the choice $]0.7, {\kern 2pt}1.3[$ just reflects the case of
a normal distribution quantile sufficiently close to unity as the
mean.

\noindent {\bf Notes:} (*) and (**) suggest further levels of
higher scale universes; (***) a continued cluster from 10$^{142}$
to 10$^{171}$, suggesting a quite different organization of matter
(in the case of (*), x varies from 82 to 120; in the case of (**),
x= 139; in the case of (***), x changes from 142 to 171).
Predictable orders of size, from the Planck scale, roughly comply
with quark-like size (10$^{10-11}$), particle to atoms
(10$^{11-17}$), molecules (10$^{21}$), human size (10$^{28}$),
stars and solar systems (10$^{40-42}$), up to the estimated upper
limit (10$^{56}$) which could be bounded by a "anti-Planck" scale
at (10$^{60-61}$).

\newcommand{\PreserveBackslash}[1]{\let\temp=\\#1\let\\=\temp}
\let\PBS=\PreserveBackslash
\begin{longtable}
{|p{24pt}|p{56pt}|p{26pt}|p{64pt}|p{45pt}|p{70pt}|p{52pt}|} a & a
& a & a & a & a & a  \kill \hline \multicolumn{3}{|p{109pt}|}
 {\begin{center} $ \ \qquad \rho = \Pi _{\rm j} \left( {\nu _{\rm j}}
\right) \qquad $ \end{center}} & \multicolumn{2}{|p{110pt}|}
  { \begin{center}
  $ \rho =$
  $\big\{ {\Pi _{\rm j} \left( {\nu _{\rm j}}  \right)
\times \left( {\Pi _{\rm j} \left( {\nu _{\rm j} - 1} \right)}
\right)} \big\} $ \end{center}}
 & \multicolumn{2}{|p{108pt}|}
 { \begin{center}
 $ \rho =$
 $\big\{ \big( {\Pi _{\rm j} ({\nu _{\rm j}}) \big)^{2} \times
\left( {\Pi _{\rm j} \left( {\nu _{\rm j} - 1} \right)} \right)}
\big\}
 $\end{center} }  \\ \hline
 \  $\rm x $
 &
    $ (1 \pm 0.3)$  $ \times 10^{{\kern 1pt}\rm x}$
 &
   \ \  $( {\nu})$
 &
   $(1 \pm 0.3)$ $\times 10^{{\kern 1pt}\rm x}$
 &
  \ \ \ $( {\nu} ) $
 &
   $(1 \pm 0.3)$ \quad $\times 10^{{\kern 1pt}\rm x}$
 &
 \ \ \  $({\nu})$
 \\
\hline
    \quad\quad \ \  0
 &
   \qquad\qquad\qquad\quad $ 1 \times 10^{{\kern 0.4pt}0} $
 &
  \quad\quad (1)
 &
  \quad\quad\qquad\qquad\qquad\qquad\qquad
  \qquad\qquad\qquad   0
 &
  \quad\quad \qquad\qquad\qquad  (1)
 &
  \quad\quad \ \qquad\qquad\qquad  \
    0
 &\\
 2
 &
 $ 1.20 \times 10^{2} $
 &
   (5)
& & & &
 \\
 10
 &
 $ 0.87 \times 10^{10}$
 &
  (14)
& & & &
 \\
  11
& & & & &
 $ 1.28 \times 10^{11} $
 &
 (7)
 \\
 17
 &
  $ 1.22 \times 10^{17}$
 &
  (19)
& & & &
 \\
  21
 &
 $ 1.12 \times 10^{21}$
 &
 (21)
& & & &
 \\
  28
 &
 $1.09 \times 10^{28}$
 &
 (27)
& & & &
 \\
29 & & &
 $ 1.27 \times 10^{29}$
 &
(17)
 & &
 \\
 31
 &
  $ 0.88 \times 10^{31}$
 &
 (29)
& & & &
 \\
 33
 &
 $ 0.87 \times 10^{33}$
 &
 (33)
& & & &
 \\
 40
 &
 $ 1.03 \times 10^{40}$
 &
(35) & & &
 $ 0.91 \times 10^{40}$
 &
(16)
\\
 42
& & &
 $ 1.26 \times 10^{42}$
 &
 (22)
& &
 \\
 56
 &
 $ 1.20 \times 10^{56}$
 &
(45)
 &
  $1.18 \times 10^{56}$
 &
(27) & &
\\
 59
& & & $ 0.93 \times 10^{59}$
 &
 (28)
&
  $ 1.33 \times 10^{59}$
 &
      (21)
 \\
 61
 &
   $ 1.24 \times 10^{61}$
& & & & &
 \\
*& *& *& *& * & * & *  \\
 82
 &
  $ 0.83 \times 10^{82}$
 &
 (60)
& & & &
 \\
 84
  & & & & &
  $ 1.29 \times 10^{84} $
&
 (27)
 \\
 99
& & &
 $ 1.12 \times 10^{99}$
& & &
 \\
 100
 &
  $ 1.20 \times 10^{100} $
 &
(70)
 & & & &
 \\
 112
& & &  $ 1.25 \times 10^{112} $
 &
 (45)
& &
 \\
 115
 &
 $ 1.13 \times 10^{115} $
& & & & &
 \\
 117
 &
  $ 0.89 \times 10^{117}$
 &
 (79)
& & & &
 \\
 120
& & & & &
  $ 1.10 \times 10^{120}$
 &
(35)
 \\
* * & * * & * * & * * & * * & * * & * * \\
 139
& & & & &
 $ 0.85 \times 10^{139}$
 &
  (39)
 \\
* * * & * * * & * * * & * * * & * * * & * * * & * * * \\
 142
 &
  $ 1.25 \times 10^{142}$
 &
 (92)
& & & &
 \\
 150
 &
  $ 1.00 \times 10^{150}$
 &
   (96)
& & & &
 \\
 163
 &
  $ 1.00 \times 10^{163} $
 &
 (103)
& & & &
 \\
 171
 &
  $ 1.24 \times 10^{171}$
 &
 (107)
 &
   $ 0.92 \times 10^{171}$
 &
  (62)
 & &
 \\
\hline
\end{longtable}

\medskip

\subsection*{4.2. About boundaries}

\hspace*{\parindent} Converging sequences of rational numbers are
known to provide the set of real numbers. However in a space of
finite dimension, real numbers cannot infer by this way. In
contrast, infinitely descending sequences (in the Mirimanoff
sense) of pairs of the $\rm {\O}^{Q}$ and $\rm {\O}^{N}$ types can
be found inside each part \{\O\}. Therefore, infinitely smaller
intervals could always be found in the lower range of scales.
These infinitely decreasing sizes are of a different nature in
that they fill each discrete part \{\O\}. However, this does not
mean that these structures necessarily are the ultimate ones.
Besides the empty set as the set with no members, one has :

\medskip \noindent
\textbf{Proposition 4.2.1.} There exists a set with no parts.

\medskip\noindent
\underline{Preliminary proof.} The space of (\O) provided with the
complementarity property gives raise to abstract sets equipotent
to sets of numbers, so that for instance ($\rm W^{m}$) owns parts
equipotent to $\rm {\mathbf{R}}^{m}$. Consider the union $\rm U =
\cup {\kern 0.7pt}(W)$ of all possible $\rm W^{m}$ spaces, and its
structural complementary in the resulting fundamental space F(U):
the structural complementary of a space with parts is a space with
no parts. Since it has no parts it cannot have members. In effect,
if its members were nonempty, this would be a nonempty space,
which is excluded by definition, while if it had empty parts, it
would just be \O. Thus, there exists a set denoted here by $(\rm
{\not{\!}{\bf c}\!\!} {\kern 2pt} {\kern 1pt})$ that has neither
members nor parts. Furthermore,  is \ $\rm \!\!\not\!{c}$
contained in none of existing sets: otherwise it would be the
complementary of Borel sets and therefore it would include parts
of itself. This provides the set of possible structures with a
lower boundary.

The fundamental set embedding space U can thus be written:
 $$
\rm F(U) =\{ \cup {\kern 1pt} (W) \} \cup {\kern 1pt} (\rm
{\not{\!}{\bf c}\!\!} {\kern 3pt}).
 \eqno(4.8)
 $$

In particular, given a partition of $\rm \cup {\kern 1pt} (W)$
into $\rm W_X$ and $\rm W_Y$, the separating distance between $\rm
W_X$ and $\rm W_Y$ in $\rm \cup {\kern 1pt} (W)$ is naught iff it
does not belong to the filter $\mathcal{F}$ holding on W, that is
since \ $\rm \!\!\not\!{c}$ and only $\rm {\not{\!}{\bf c}\!\!}
{\kern 2pt} {\kern 1pt} \in {\mathcal{F}}$:
 $$
\rm \Delta_{{\kern 1pt}\cup {\kern 1pt}(W)}(W_X,{\kern 1pt}W_Y)
=(\rm {\not{\!}{\bf c}\!\!} {\kern 2pt} {\kern 1pt}).
 \eqno(4.9)
 $$

The empty hyperset can no longer be treated as it was formerly
considered the well founded empty set: in particular, ${\O}\notin
\mathcal{F}$ whereas ${\O}^{{\kern 0.3pt}\O}$ has not the same
status.

\medskip\noindent
\textbf{Corollary 4.2.1.} The set with neither member nor parts is
strictly unique.

\medskip\noindent
\underline{Proof.} Let two universes $\rm F(Ui) = \{\cup {\kern
2pt}(Wi) \} \cup ({\not{\!}{\bf c}\!\!} {\kern 2pt}{\kern 2pt}{\bf
i})$ and $\rm F(Uj) = \{ \cup {\kern 2pt}(Wj)\} \cup
({\not{\!}{\bf c}\!\!} {\kern 2pt}{\kern 2pt}{\bf j})$ as in
(3.8). Then, $({\not{\!}{\bf c}\!\!} {\kern 2pt}{\kern 2pt}{\bf
i}) \cup  ({\not{\!}{\bf c}\!\!} {\kern 2pt}{\kern 2pt}{\bf j}) =
({\not{\!}{\bf c}\!\!}{\kern 2pt}{\kern 2pt})$ since the reunion
of no parts and no member is just identical with (0). However this
would also allow one to write: $({\not{\!}{\bf c}\!\!}{\kern
2pt}{\kern 2pt}) = \{ ({\not{\!}{\bf c}\!\!} {\kern 2pt}{\kern
2pt}{\bf j}), {\kern 1pt} ({\not{\!}{\bf c}\!\!} {\kern 2pt}{\kern
2pt}{\bf j})\}$ and $({\not{\!}{\bf c}\!\!} {\kern 2pt}{\kern
2pt}{\bf j})$ would be composed of two parts, which is
contradiction.

Set $({\not{\!}{\bf c}\!\!}{\kern 2pt}{\kern 2pt})$ can thus be
denoted as the "nothingness singleton" \{${\not{\!}{\bf
c}\!\!}{\kern 2pt}{\kern 2pt}$\}.

\medskip\noindent
\textbf{Corollary 4.2.2.} A set equipotent to the set of natural
integers {\bf N} is intrinsically of nonzero measure, and a
segment E of {\bf N} cannot be of measure naught even relatively
to the correspondingly available segment of {\bf Q}. A member of
$\rm \mathbf{N}^{\mathbf {x}}$ is never of measure naught.

\medskip\noindent
\underline{Preliminary proof.} (i) given {\bf N} and only {\bf N}
as the fundamental set, one cannot insert each member in an
interval as small as needed, since there exists no segment
available with size lower than 2 units, able to contain each
point. For this to be achieved it is necessary to provide the
system with at least $\mathbf{N} \times \mathbf{N}$ so as to
generate {\bf Q}. The former case is called the intrinsic measure
on {\bf N}, while the latter is the measure on {\bf N} relatively
to {\bf Q}.

\medskip \noindent
(ii) Let E(\O) be equipotent to a beginning section of ({\bf N}):
the result presented above states that there exists a finite
number of rational inferring from the cartesian product of this
space. Therefore it is not possible to insert the members of E in
a sum of intervals of $\rm E \times E$ as small as needed. Thus
E(\O) can be of neither intrinsic nor {\bf Q}-relative measure
naught.

\medskip \noindent
(iii) Let finally $\rm E^{x} = (\mathbf {N} \times \mathbf{N}
\times ... \times \mathbf{N})_{{\kern 1pt} finitely  {\kern 2pt} x
{\kern 2pt} times, {\kern 2pt} x<1}$: each member of the
equipotent set is an ordered pair. Even if an unordered N-uple
\{a,b, ...\}, whatever the nature of a and b, could be eventually
of measure zero, excepted intrinsically, an ordered N-uple (ab...)
owns a dimension $\geq1$ and cannot be inserted in an interval as
small as needed. \qquad\qquad\qquad\qquad\qquad\qquad
\qquad\qquad\qquad \qquad\qquad\qquad  \ \ \ \ (Q.E.D.)

\medskip\noindent
\textbf{Corollary 4.2.3.} A finite set is not of measure zero if
its contains members and parts that are representative of a space
of dimension $\rm D > 1$.

\medskip\noindent
\textbf{Corollary 4.2.4.} Due to uniqueness of $({\not{\!}{\bf
c}\!\!}{\kern 2pt}{\kern 2pt})$, the "tessellattice" is correctly
tessellated since no gap can subsist between any two or more of
its empty tessellation balls. Furthermore, ${\not{\!}c\!\!}{\kern
2pt}{\kern 2pt}$ provides the tessellattice with a infinum, and
thus with a partial order.

\section*{5. Particles in a lattice universe}

\subsection*{5.1. Introduction}

\hspace*{\parindent} Let space be represented by the lattice $\rm
F(U) = \cup {\kern 2pt}(W) {\kern 1pt} \cup {\kern 1pt}
{\not{\!}c\!\!}{\kern 2pt}{\kern 2pt}$ {\kern 1pt} as from
relation (9.1) in Bounias and Krasnoholovets (2002), where
${\not{\!}c\!\!}$ {\kern 1pt} is the set with neither members nor
parts. This accounts for both relativistic space and quantic void,
since: (i) the concept of distance and the concept of time have
been defined on it, and (ii) this space holds for a quantum void
since on one hand, it provides a discrete topology, with quantum
scales, and on the other hand it contains no "solid" object that
would stand for a given provision of physical matter.

The above relation (3.2) involves the mapping of a frame of
reference into its image frame of reference in the next section of
space-time. Without such a continuity there would be no
possibility of assessing the motion of any object in the perceived
universe. This is exactly a case of "analysis situs", in the
original meaning used by Poincar\'{e}. Now, continuity in the
perception of a space-time is provided iff the frames of
references are conserved through homeomorphic mappings. This means
that there is no need for exact replication: just topological
structures should be conserved. Hence, the realization of
varieties if allowed, even in a space of different dimension. This
supports the following:

\medskip\noindent
\textbf{Proposition 5.} The sequence of mappings of one into
another structure of reference (e.g. elementary cells) represents
an oscillation of any cell volume along the arrow of physical
time.

However, there is a case in which a threshold may exist,
precluding the conservation of homeomorphisms: let a
transformation of a cell involving some iterated internal
similarity (see Figure 3 for simplified example). Then, if N
similar figures with similarity ratios 1/r are obtained, the
Bouligand exponent (e) is given by
 $$
\rm N{\kern 1pt}(1/r)^{{\kern 1pt}e}=1
 \eqno(5.1)
 $$
and the image cell gets a dimensional change from d to $\rm
d^{{\kern 0.4pt}\prime} = \ln(N)/\ln(r) = e>l$.

Then, the putatively homeomorphic part of the image cell is no longer a
continued figure and the transformed cell no longer owns the property of a
reference cell.

This transformation stands for the formation of a "particle" also
called "particled cell" or more appropriately "particled ball",
since it is a kind of topological ball $\rm B[{\O},{\kern
1pt}r({\O})]$. Thus the following:

\begin{figure}
\begin{center}
\includegraphics[scale=1.7]{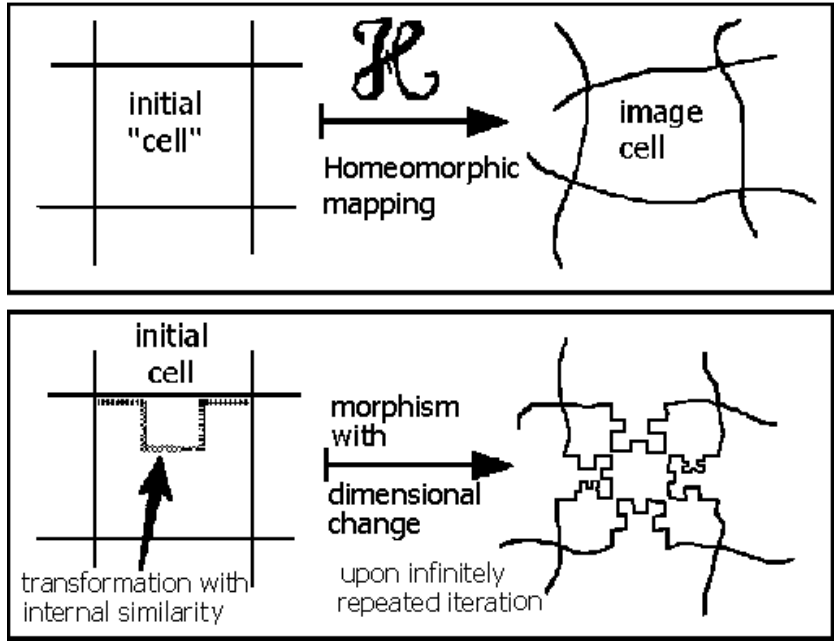}
\caption{\small{The continuity of homeomorphic mappings of
structures is broken if once a deformation involves an iterated
transformation with internal self-similarity, which involves a
change in the dimension of the mapped structure. Here, various the
first 2 or 3 steps of the iteration are sketched, with basically
the new figure jumping from (D) to approximately ($\rm D + 1.45$).
The mediator of transformations is provided in all cases by empty
set units.}} \label{Figure 3}
\end{center}
\end{figure}

\medskip\noindent
\textbf{Statement 5.} A particled ball is represented by a
nonhomeomorphic transformation in a continuous deformation of
space elementary cells.

\subsection*{5.2. On quanta of fractality and fractal
decomposition}

\hspace*{\parindent} Before examining the interactions of
particled balls with the degenerate space lattice and further with
other particled balls, it is necessary to demonstrate some
mathematical preliminaries.

\subsubsection*{5.2.1. Quanta of fractality}

\hspace*{\parindent} A minimum fractal structure is provided by a
self-similar figure whose combination rule includes a initiator
and a generator, and for which the similarity dimension exponent
is higher than unity.

\medskip \noindent
(i) \textbf{Initiator.} Due to self-similarity of $\geq {\O}$,
each time one considers the complementary of itself in itself, one
gains: ${\O} \mapsto \{ ({\O}),{\kern 1pt} {\O}\}$. That is, one
ball gives two identical balls. This is continued into a sequence
of \{1/2, 1/4, ..., 1/2n\} numbers at the n$\rm ^{th}$ iteration.
Thus, the series $\rm (I) = \sum_{{\kern 1pt} i=1\rightarrow
\infty} \{1/2^{{\kern 1.5pt}i}\}$ stands for the initiator,
providing the needed iteration process. The terms of (I) are
indexed on the set of natural numbers, and thus provide an
infinitely countable number of members.

Interestingly, $\rm 2^{{\kern 1pt}n}$ also denotes the number of
parts from a set of n members,

\medskip \noindent
(ii) \textbf{Generator.} Let an initial figure (A) be subdivided
into r subfigures at the first iteration. The similarity ration is
thus $\rm \varrho = 1/r$. Let $\rm N=(r+a)$ be the number of
subfigures constructed on the original one. Then, one has $\rm e =
Ln{\kern 1pt}(r+a) / Ln{\kern 1pt} N$. The value of e is bounded
by unity if r is extended to infinity. For any r finite (likely
the case in a physical world), the exponent e is above unity.
Then:
 $$
\rm \{ \min {\kern 1pt}(e) |{\kern 2pt} e>1 \} = Ln \big(
\max{\kern 1pt} (r) + 1 \big) /{\kern 1pt} Ln \big( \max (r)
\big).
 \eqno(5.2)
 $$
This completes the description of a quantum of fractality.

\subsubsection*{5.2.2. The fractal decomposition principle}

\hspace*{\parindent} Let a fractal system be denoted by $\Gamma$,
such as $\rm \Gamma = \{ ({\O}), {\kern 1pt}(r+a)\}$. More complex
systems just need to be incorporated several different subfigures
to which the following reasoning could be extended. At the $\rm
n^{th}$ iteration, the number of additional subfigures is $\rm N_n
= (r+a)^{n}$. The similarity ration becomes $\rm \varrho _n =
l/r^{{\kern 1pt}n}$. Owing to subvolumes ($\rm v_i$ constituted at
each iteration, in the simplest case $\rm v_i = v_{i-1}{\kern
1pt}(1/r)^{{\kern 1pt}3}$. Since at the i$^{\rm th}$ iteration as
many as $\rm N_i = (r+a)^{{\kern 1pt}i}$ such subvolumes are
created, the total volume occupied by the subvolumes formed by the
fractal iteration to infinity is the sum of the series:
 $$
\rm v_{f} = \sum_{(i=1 \rightarrow \infty)} \{(r+a)^{{\kern 1pt}
i} \cdot v_{i-1}{\kern 1pt}(1/r)^{{\kern 1pt}3} \} \quad\quad
 \eqno(5.3{\kern 0.3pt} \rm a)
 $$
that can be developed into
 $$
\rm v_{f} = \sum_{(i=1 \rightarrow \infty) } \{ [\Pi _{(i=1
\rightarrow n}]  (r+a)^{i-1}/ (r)^{3{\kern 0.5pt}i} \}.
 \eqno(5.3{\kern 0.3pt} \rm b)
 $$

This leads to the following definition:

\medskip \noindent
\textbf{Definition 5.2.} A fractal decomposition consists in the
distribution of the members of the set of fractal subfigures:
  $$\rm
\Gamma \supset  \big\{ \sum_{{\kern 1pt} i=1 \rightarrow \infty}
\{(r+a)^{{\kern 1pt}i} \cdot v_{i-1} {\kern 1pt}(l/r)^{3} \}
\big\}
  $$
constructed on one figure, among a number of connected figures
($\rm C_1, {\kern 2pt}C_2,{\kern 2pt} ...,{\kern 2pt} C_k$)
similar to the initial figure (A). If k reaches infinity, then all
subfigures of A are distributed and (A) is no longer a fractal.

\begin{figure}
\begin{center}
\includegraphics[scale=1.6]{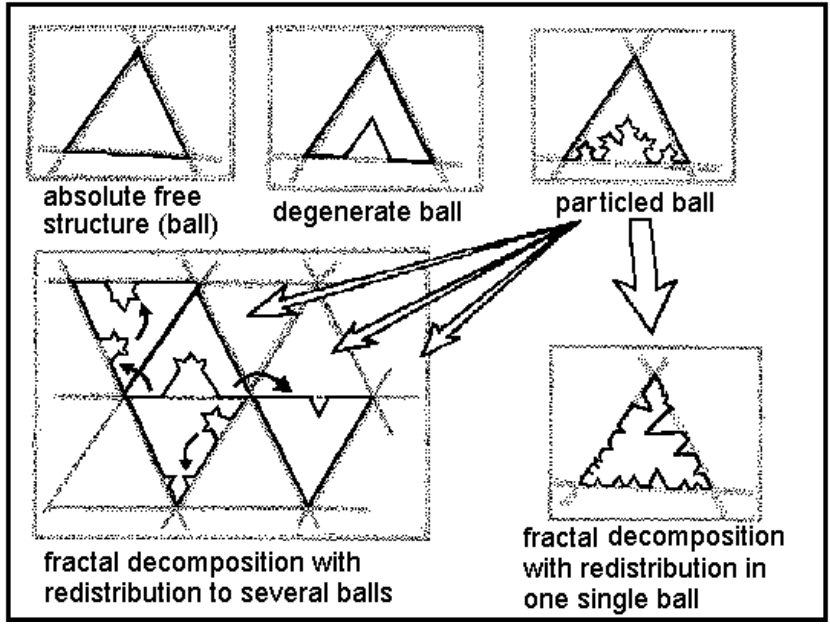}
\caption{\small{A canonical ball is represented as a triangle,
figuring 3 dimensions, in a metaphorical form. A degenerate ball
keeps the same dimension, in contrast with a particled ball
endowed with a fractal substructure. A complete decomposition into
one single ball ($\rm k=1$) conserves the volume without keeping
the fractal dimension. The von Koch-like fractal has been
simplified to 3 iterates for clarity.}} \label{Figure 4}
\end{center}
\end{figure}

Figure 4 shows in a very schematic way the fractal decomposition
process.

\medskip

\noindent \textbf{Corollary 5.2.} Reciprocally, a fractal figure
can be recomposed from an infinitely enumerable set of
self-similar figures whose numbers and sizes are distributed as in
relation (5.2)-(5.2a).

\subsection*{5.3. Interactions involving exchanges of structures}

\subsubsection*{5.3.1. Scattering from friction}

\hspace*{\parindent} Let a ball (A) containing a fractal subpart
on it, as shown in Figures 3 and 4. Deformations can be
transferred from one to another ball with conservation of the
total volume of the full lattice (which is constituted by a higher
scale empty set). If a fractal deformation is subjected to motion,
it will collide with surrounding degenerate balls. Such collisions
will result in fractal decompositions at the expense of (A) whose
exponent (e$_{\rm A}$) will decrease, and to the profit of
degenerate cells: if k is finite, one will have $\rm (e_{{\kern
1pt} i})>l,{\kern 2pt} (e_{{\kern 1pt}2})>l, {\kern 1pt} ...,
{\kern 1pt}(e_{{\kern 1pt}k})> 1$. A fractal decomposition gives
raise to a distribution of coefficients $\rm {\it f}(e_{{\kern
1pt}k})$, whose most ordered form is a sequence of decreasing
values:
 $$
\rm {\it f}(e_{{\kern 1pt}k}) = \{ (e_{{\kern 1pt} i})_{{\kern
1pt}(i {\kern 1pt} \in {\kern 1pt} ]k, {\kern 1pt} 1 ])} \}.
 \eqno(5.4)
 $$

\subsubsection*{5.3.2. Boundary conditions}

\hspace*{\parindent} From relation (5.4), it follows that the
remaining of fractality decreases from the kernel (i.e. the area
adjacent to the original particled deformation) to the edge of the
inerton cloud. At the edge, it can be conjectured that, depending
on the local resistance of the lattice, the last decomposition
(denoted as the $\rm n^{th}$ iteration) can result in ($\rm
e_{{\kern 1pt}n}) = 1$. Since in all cases one has ($\rm e_{{\kern
1pt} n-1} >l$, eventhough the corresponding remaining deformation
is a fragment of the original fractal structure, then the
resulting non-fractal deformations can be theoretically
distributed up to infinite distance.

Therefore, while central inertons exhibit decreasing higher
boundaries, edge inertons are bounded by a rupture of the
remaining fractality.

\subsection*{5.4. Discussion and alternative hypothesis}

\subsubsection*{5.4.1. About infinitesimal elements}

\hspace*{\parindent} One could wonder whether an infinite
iteration could not physically take a infinite time and thus would
be non-physical, (i) One answer is that all from the first
iteration would take a time decreasing to infinitely small values
to occur, (ii) Another would state that a quantum of fractality
with the simplest geometrical generator and initiator would also
stand for a quantum of time corresponding with the maximum
velocity for the corresponding particle, and that the velocity of
a non-fractal deformation would stand for the maximum velocity of
non-massive corpuscles.

\subsubsection*{5.4.2. Incomplete fractality hypothesis}

\hspace*{\parindent} However, one could also conjecture that mass
could be in some way proportional to the number of iterations on
the way to a fractal whose completion would be only a theoretical
(likely hyperbolic) limit. Here again, the number of iterations
provide an alternative kind of qualitative jump, where $\rm n=1$
iteration would stand for non-massive corpuscle, and $\rm n>2$ for
massive particles.

\subsubsection*{5.4.3. Topological alternatives}

\hspace*{\parindent} Some other features about massive versus
non-massive properties remain to be explored. Among other
candidates, one could list dense versus non-dense, compact versus
not compact, complete versus not complete subspaces.
Conjecturally, one could envision that the space of particles
could be everywhere dense in that of superparticles while
non-dense in the total space, and the space of superparticles
could be dense in the total space. This is matter of work in
progress.

\section*{6. Practical predictions from the inertons theory}

\subsection*{6.1. Preliminaries}

\hspace*{\parindent} A particled ball as described above provides
a formalism describing the elementary particles proposed by
Krasnoholovets and Ivanovsky (1993), Krasnoholovets (1997; 2000).
In this respect, mass is represented by a fractal reduction of
volume of a ball, while just a reduction of volume as in
degenerate cells is not sufficient to provide mass. Accordingly,
if $\rm v_o$ is the volume of an absolutely free cell, then the
reduction of volume resulting from a fractal concavity is the
following: $\rm V^{part} = v_o - v_f$, that is, according to
relation (5.3b):
 $$
\rm V^{part}=v_o \cdot \Big( 1 - \sum_{(i=1 \rightarrow \infty)}
\{ [ {\kern 1pt} \Pi_{(i=1 \rightarrow n)} {\kern 1pt}
(r+a)^{{\kern 1pt} i-1}]{\kern 1pt} / {\kern 1pt} (r)^{{\kern 1pt}
3{\kern 0.7pt}i} \} \Big),
 \eqno(6.1{\kern 0.3pt}\rm a)
 $$
\noindent that is, since $\rm (r+a) = (r)^{{\kern 1pt}e}$, we have
instead of (6.1a)
 $$
\rm V^{part}=v_o \cdot \Big( 1-\sum_{(\nu)}\big( \sum_{(i=1
\rightarrow\infty)} \{ [ {\kern 1pt} \Pi_{(i=1 \rightarrow n)}
{\kern 1pt} (r_\nu)^{{\kern 1pt}e_\nu (i-1)} ]  {\kern 1pt} /
{\kern 1pt} (r_\nu)^{{\kern 1pt}3{\kern 0.7pt}i} \} \big)_{\nu}
\Big)
 \eqno(6.1{\kern 0.3pt}\rm b)
 $$
where ($\nu$) denotes the possibly several fractal concavities
affecting the particled ball. This point will be the matter of an
examination of the various kinds of particles predicted by the
model in Part 3 of this study.

Relation (6.1b) relates the volume of particled balls to the
fractal dimensional change (e), which can be expressed as the
following:

\medskip
\noindent \textbf{Proposition 6.1.} The mass Ma of a particled
ball A is a function of the fractal-related decrease of the volume
of the ball:
 $$
\rm Ma \varpropto (1/{\kern 1pt}V^{part}) \cdot (e_{{\kern
0.5pt}v}-1)_{{\kern 1pt}e_{{\kern 0.5pt}v}>1}
 \eqno(6.2)
 $$
where (e) is the Bouligand exponent, and $\rm (e-1)$ the gain in
dimensionality given by the fractal iteration.

Just a volume decrease is not sufficient for providing a ball with mass,
since a dimensional increase is a necessary condition.

\subsection*{6.2. Foundations of the inertons theory}

\hspace*{\parindent} Two interaction phenomena are considered:
first, the elasticity ($\gamma$) of the lattice favors an exchange
of fragments of the fractal structure between the particled ball
and the surrounding degenerate balls. In a first approach, the
resulting oscillation has been considered homogeneous. Second, if
the particled ball has been given a velocity, its fractal
deformations collide with neighbor degenerate balls and exchanges
of fractal fragments occur.

\medskip
\noindent \textbf{Proposition 6.2.} The velocity of the transfer
of deformations is faster for non fractal deformations and slower
for fractal ones, at slowering rates varying as the residual
fractal exponent ($\rm e_{{\kern 1pt} i}$).

\medskip
\noindent \underline{Justifications.} A fractal subvolume owns an
infinite surface, which imposes a more important transfer than the
progression of a non fractal volume, which involves a finite
surface. More generally each iteration step involves a proper
quantum of transfer time.

\medskip\noindent
\textbf{Proposition 6.3.} The motion of the system constituted by
a particled ball and its inerton cloud provides the basis for de
Broglie and Compton wavelengthes.

\medskip \noindent
\underline{Justifications.} During the progression of a particled
ball, its mass is progressively transferred to a cloud of
surrounding balls which get fragments of the particle mass. These
new quasi-particles are called "inertons". The inertons velocity
is faster due to lower fractal dimension, and the cloud migrates
forwards up to the state where the residual mass of the particle
is low enough. At this step, the particle has progresively lost
its velocity due to the collisions: then, since collision with the
degenerate lattice balls stop, no more inertons are produced. At
this time, the elasticity of the lattice starts to reinject the
fragments of deformations and progressively restore the initial
fractal in the particled ball, which is reconstructed. Similarly,
the equivalent in momentum (that is ($\langle m \rangle
\!\!\cdot\!\! \langle v \rangle $), where $\langle m \rangle$ and
$\langle v \rangle$ respectively denote the average mass and
velocity of the system composed with \{particle + inertons\} and
further denoted ($\Xi$) is progressively retransfered to the
particled ball, which gets back its initial velocity.

Then, the trajectory of a particle is represented by a complex
oscillating system which drives from a state (denoted as initial)
where the particle owns its full mass and velocity to a state
where its velocity is minimal (eventually null) and part of its
mass has been transferred to the inerton cloud (Table 2).

This represents a cycle with a period $\lambda$ of the path, such
that \textsf{min}$\rm \{M, {\kern 2pt}V^{part}\}$ occurs at
$\lambda/2${\kern 0.5pt}: this defines a parameter identified with
the de Broglie wavelength. Then, the system ($\Xi$) which exhibits
similarities with a crystallite has an oscillating size called
$\widetilde \lambda_{{\kern 1pt}v_{0}}$ and identified with the
Compton wavelength (Krasnoholovets, 1997, 2000, 2001). The
inertons cloud itself is characterized by a specific amplitude
$\Lambda$ that has been related with the other two ones by the
following relations:
 $$
\Lambda = \lambda {\kern 2pt} \hat{c}/v_{\rm o} = \widetilde
\lambda_{{\kern 1pt}v_{\rm o}} \cdot {\hat c}^{{\kern
0.3pt}2}/v^{\kern 1pt 2}_{\rm o}
 \eqno(6.3)
 $$
where $v_{\rm o}$ is the initial velocity of the particled ball
and $\hat c$ is the velocity of emitted inertons.

Table 2 summarizes the main steps and features of the particle to
inertons cycle.

\medskip
\noindent \textbf{Remark 6.1.} The center of mass of the system
($\Xi$) permanently oscillates between the original particle and
the inerton cloud. This introduces an additional parameter whose
various forms provide a solid support to the concept of spin
(Krasnoholovets, 2000).

\medskip
\noindent \textbf{Remark 6.2.} The system composed with the
particle and its inertons cloud is not likely to be of homogeneous
shape. Therefore, their relative motion will expectably exhibit an
eddy-like form (Wu and Lin, 2002). This property will be accounted
for in spin-related properties of observable matter.

\medskip
\noindent \textbf{Proposition 6.4.} The fractality of
particle-giving deformations gathers its space parameters $\rm
(\varphi {\kern 1pt} i){\kern 1pt}i$ {\kern 1pt} and velocities
($\mathfrak{v}$) into a self-similarity expression, which provides
a space-to-time connection.

\medskip \noindent
\underline{Justifications.} Let $\rm (\varphi _o)$ and
$\mathfrak{v}_{\rm o}$ be the reference values. Then the
similarity ratios are $\rm \varrho (\varphi) = (\varphi {\kern
0.6pt}i/ \varphi _o)$ and $\varrho (\mathfrak{v}) = (\mathfrak{v}
/ \mathfrak{v}_{\rm o})$. Therefore
 $$
\rm \varrho(\varphi)^{{\kern 1pt}e} + \varrho
({\mathfrak{v}})^{{\kern 1pt} e} = 1.
 \eqno(6.4 {\kern 0.3pt} \rm a)
 $$

Since  $\rm (\varphi {\kern 1pt} i){\kern 1pt}i = {\kern 1pt}
\big\{ ${\kern 1pt}\textsf{distances} (L) and \textsf{masses}
($\mathfrak{m})\big\}$, one can write: $\rm \mathfrak{m}_{{\kern
0.3pt}o}/\mathfrak{m} = L/L_{{\kern 0.3pt}o}$, so that
 $$
\rm (\mathfrak{m}_{{\kern 0.3pt} o}/\mathfrak{m})^{{\kern 1pt}e} +
({\mathfrak{v}}/{\mathfrak{v}}_{{\kern 1pt} o})^{{\kern 1pt} e} =
1.
 \eqno(6.4 {\kern 0.3pt} \rm b)
 $$

While coefficient (e) gets a value above unity, the geometry
outside $\rm e=1$ escapes the usual $\rm (3-D{\kern 1pt} + {\kern
1pt}t)$ space-time and, owing to the previously demonstrated
necessity of an embedding 4-D (timeles) space, the coefficient (e)
must reach $\rm e=2$. Hence, the boundary conditions provide the
following results:
 $$
\rm (\mathfrak{m}_{{\kern 0.3pt}o}/\mathfrak{m})^{2} +
({\mathfrak{v}}/{\mathfrak{v}}_{{\kern 0.3pt}o})^{2} = 1 \
\Leftrightarrow \ \mathfrak{m} = \mathfrak{m}_{{\kern 0.3pt}o} /
\sqrt{1 - ({\mathfrak{v}}/{\mathfrak{v}}_{{\kern 0.3pt}o})^{2}}
 \eqno(6.4{\kern 0.3pt} \rm c),
 $$
 $$
\rm (L/L_{{\kern 0.3pt} o})^{2} +
({\mathfrak{v}}/{\mathfrak{v}}_{{\kern 0.3pt} o})^{2} = 1 \
\Leftrightarrow \ L = L_{{\kern 0.3pt}o}\cdot \sqrt{l -
({\mathfrak{v}}/{\mathfrak{v}}_{{\kern 0.3pt}o})^{2}}.
 \eqno(6.4{\kern 0.3pt}\rm d)
 $$

\medskip

\textbf{Table} 2. The sequence of events occurring during a cycle
of the behavior of the system composed of a moving particled
deformation and its inertons cloud. \break {\bf Note:} The
velocity of transmission of the deformations from the particled
ball to the degenerate balls is a nonlinear function of the
following variables: $\rm V=F(\gamma,{\kern 2pt} K, {\kern 2pt}
\tau, {\kern 2pt}\omega, {\kern 2pt}R; {\it f}(e))$ where $\gamma$
is the elasticity factor;  K is the tenseness of the lattice;
$\tau$ is the transmittivity of deformations; $\omega$ is the
resistance of the lattice; R is the reaction to the emission of
deformations; $f(\rm e)$ is the fractal characteristic.

\begin{longtable}
{|p{42pt}|p{70pt}|p{64pt}|p{69pt}|p{64pt}|p{52pt}|} a & a & a & a
& a & a  \kill
 \hline

\bigskip \qquad \qquad\qquad\qquad \qquad\qquad\quad\quad Phase
state
\par
 &

\bigskip \qquad \qquad\qquad\qquad \qquad\qquad\quad
 {Start point}
\par &
\[
 ]0,\,\,\lambda /2 [
\]
 \par
  &

\bigskip
\qquad\qquad
\[
\lambda /2
\]
&
\[
]\lambda /2,\,\,\lambda [
\]
 &
 \begin{center}
$ \lambda $
 \end{center}
 \\
\hline
 particled ball \par
 &
 \footnotesize{max velocity $v_{\rm o}$; \par max mass $\rm M_o$; \par max
 tenseness; \par max fractal (e) }\par
 &
\footnotesize{mass loss; \par velocity loss; \par tenseness de-\ \
\  \break  crease}
\par
 &
\footnotesize{min mass; \par min velocity; \par min tenseness;
\par balance with \ \ \break cloud}
\par
 &
\footnotesize{reincrease of \break mass and velocity} \par
 &
\footnotesize{return to \break initial state with max \break mass
and \break velocity}
\par  \\

inertons cloud
 &
\footnotesize{min mass $\rm m_o$=0;  \par not yet resistance to
the motion of particled deformation}
 &
\footnotesize{collisions with lattice: emission of inertons with
high speed (low lattice tenseness)}
\par
 &
\footnotesize{max mass; \par max reaction; \par
 max tenseness \break as opposed to \break
 particle motion; \par
 max dispersion of inertons by the degenerate
 lattice}
 &
\footnotesize{reaction decre-\break ase; \par return of mass;}
 \par
\footnotesize{decrease of \ \break tenseness}
 &
\footnotesize{dissappea- \break rance of \ \  \break  the inerton
cloud}
\\


\quad\quad\quad \ \ \ {\kern 1pt}{\kern 1pt}{\kern 1pt} whole
\break  system
 &
  \qquad\qquad\qquad\qquad
\footnotesize{no inerton  re-\break  action;}  \par

\footnotesize{resistance from degenerate lattice;}
  \par
 &
  \qquad\qquad\qquad\qquad
\footnotesize{tenseness ba-\break lance and fle-\break xibility of
lat-\break tice
  inertons  \ \ \break reaction}
 &
 \qquad\qquad\qquad\qquad
\footnotesize{balance in the \break respective tenseness and \ \
\break center of mass}
 \qquad
 &
 \qquad\qquad\qquad\qquad
\footnotesize{reverse change in localization of the center of
mass}

 &
 \qquad\qquad\qquad\qquad\qquad
\footnotesize{de Broglie \ \break period reached} \\

\hline

\end{longtable}

\medskip
\noindent \textbf{Remark 6.3.} The Lagrangian $(\mathcal{L})$
should obey a similar law and $(\mathcal{L}/\mathcal{L}_{{\kern
0.3pt} \rm o})$ should fulfill relation (6.4b) as a form of
$\varrho{\kern 1pt}(\varphi)^{{\kern 0.3pt}\rm e}$. Then, $\rm
(\mathcal{L}/\mathcal{L}_{{\kern 0.3pt} \rm o})^{{\kern 0.7pt}e} +
(\mathfrak{v}/\mathfrak{v}_{{\kern 0.3pt}\rm o})^{{\kern 0.7pt}2}
= 1$ and analogically one could take $\rm \mathcal{L}_{{\kern
0.3pt}o} = - {\mathfrak{m}} \mathfrak{v}_{{\kern 0.3pt}\rm
o}^{{\kern 1pt}2}$. Thus finally $\mathcal{L}=- \mathfrak{m}
\mathfrak{v}_{{\kern 0.3pt} \rm o }^{{\kern 1pt}2} \sqrt{1 -
(\mathfrak{v}/\mathfrak{v}_{{\kern 0.3pt}\rm o})^{{\kern 1pt}2}}$.

By analogy with special relativity, $\mathcal{L},{\kern 2pt}
\mathfrak{m}, {\kern 2pt}\mathfrak{v}$ are the parameters of a
moving object, while $\mathfrak{v}_{{\kern 0.3pt} \rm o} = c$
where $c$ is the celerity of light. This supports some
requirements pointed by Krasnoholovets (2001).

\subsection*{6.3. Wave mechanics analysis}

\hspace*{\parindent} Work (Krasnoholovets, 2002) has provided the
derivation of wave equations from the inerton system. In short,
given a set $\{\pi\}$ of two parameters describing the behavior of
respectively the mass of the particle and the total inertons cloud
(called the "rugosity" of the surrounding space, due to the
distribution of scattered fragments of fractal deformations), the
equations take the following general form:
 $$
{\ddot \pi} - {c}^2_{\pi} \nabla \pi = 0
 \eqno(6.5)
 $$

Solutions of (6.5) provide a real macroscopic wave function,
allowing to find back the equivalent of the Schr\"odinger
equation.

The same works finally derive also a mass field accounting for
gravitation, and proves that the inert mass and the gravitational
mass are the same. Hence, cosmic scale properties are inferring
from particle scale characteristics.

In Part 3 of this work, further corollaries will be derived about
the origin and classification of families of particled balls, and
in connection with the former, it will be deduced an explanation
on the mechanics underlying the origins of our observable
space-time.

To which extent the motion of the \{particle + inertons\} system
should follow the eddy-like motion that nonhomogeneous systems
should follow (Lin and Wu, 1998) is matter of further
investigations.

\section*{7. Experimental assessment of the inertons \break existence}

\hspace*{\parindent} Some preliminary experimental verifications
have already been achieved, and some protocols for further proofs
can be proposed.

\subsection*{7.1. Former evidence}

\hspace*{\parindent} The prediction of collective behaviour of
atoms in solid matter from the existence of the \{particle +
inertons\} system has been tested on both physical and chemical
systems.

\medskip
\noindent \textbf{7.1.1.} Moving electrons emit inerton clouds
which can be detected in the form of anomalous photoelectric
effects (Krasnoholovets, 2001{\kern 1pt}b).

\medskip
\noindent \textbf{7.1.2.}  The impact of inertons on the
collective behaviour of atoms in various metals has been evaluated
and then experimentally observed by high resolution electron
microscopy scanning (Krasnoholovets, and Byckov, 2000).

\medskip
\noindent \textbf{7.1.3.} The existence of inerton clouds has been
calculated for hydrogen atoms clustering in the
$\delta$-KH(IO$_3$)$_2$ crystal and the proton dynamic study has
verified the theory (Krasnoholovets, 2001{\kern 1pt}d).

\subsection*{7.2. Further experimental protocols}

\hspace*{\parindent} Former experiments have confronted the
recording of signals from informational fields other than
electromagnetic ones, like the Kozyrev effect and others (Kozyrev
and Nazonov, 1978). Now, Krasnoholovets et al. (2001) have
proposed a series of protocols for testing the emission of
"inerton radiation" in various conditions. The measurements will
be performed by using pyroelectric sensors constructed for this
purpose. The project includes the following three cases:

\medskip \noindent
\textbf{7.2.1.} Prediction of inerton emission by the sun, and
observation of fluctuations of their field.

\medskip \noindent
\textbf{7.2.2.} Prediction and measurement of the velocity of
inerton waves emitted from distant stars whose parameters are
sufficiently well known.

\medskip \noindent
\textbf{ 7.2.3.} Prediction and measurement of inerton flows
emitted from satellites of solar planets.

\section*{8. Discussion and conclusions}

\hspace*{\parindent} The most important problem faced by most of
physical theories is that they pose some assumptions which are
founded on observations and speculations whose validity can hardly
be supported excepted by indirect testing. About speculations, for
instance classical theories pose the existence of elementary
particles, string theory supposes the existence of free strings
and of masses whose linear density gives the tension of the
strings (Maldacena, 2000), while the rolling up of a membrane into
a cylinder of infinitesimal diameter has been identified with a
1-D line (Duff, 1998), which is topologically wrong; lower
boundaries are hypothesized for a Hamiltonian, if any, to account
for a still ill-defined void state, while upper limits are
required for the number of quantum states which are not postulated
but resulting from statistical developments, and should be
smoothed at large scales so as to let a discrete system be
replaced by a continuous limit ('t Hooft, 1999). The latter point
implicitly suggests that the observable universe is infinite,
which remains an open question.

At observational levels, it remains difficult to get decisive
answers for essentially two reasons. First, experimental data can
often be interpreted in several ways: de Broglie noted that the
same mathematical equations can have several [physical]
explanations (Rothwarf, 1998), and Maldacena (2000) points that an
inconsistent theory could agree with experiment. Second, even the
measures of physical phenomena returns uncertain data: while the
homogeneity and the isotropy of universe is required around each
point for applicability of the FRW metrics (Smoller and Temple),
the uniformity od universes appears as paradoxical (Bucher and
Spergel, 1999) and the claim for anisotropy-supporting data
(Ralston, Jain and Nodland, 1997) consistently raises polemic
reactions (Ralston and Nodland, 1997).

Between these extreme positions, there unfortunately lies the
situation in which no appropriate measure can decide between
alternative hypothesis: this is the case for whether the expansion
works for nearby galaxies only, or for the whole universe (Smoller
and Temple, 2000), and what would be expanding between two
independent objects (Walker, 1996; Bucher and Spergel, 1999). That
objects can be independent, and why the expansion of "space" would
not affect at all the microscopic world remain as many postulates,
since the measuring devices may not be independent from the
measured phenomena.

Our approach aimed at trying to pose as few postulates as
possible, and to rather examine which kind of probationary
space(s) and mathematical properties would fulfill the conditions
required to support a proposition such as: there exists a physical
universe embedding a self-perceived phenomenon that we use to call
life. This drove to the identification of a primary axiom as the
existence of the empty set, in turn providing existence of
abstract mathematical spaces. Then, spaces of topologically closed
objects gives raise to physical-like spaces, up to the function of
self-conscious perception (Bounias, 2000). Several main aspects of
this model will now be examined with respect to usual requirements
for space-time structures and properties.

\medskip
\noindent \textbf{8.1. Space-time continuity and quantum
structures}

\medskip

Such a physical-like space is therefore composed of discrete cells endowed
with quantumwise defined relative scales and whose interior is potentially
provided with the power of continuum. An important consequence is that this
property lays a bridge exactly on the gap separating the so-far discrete
nature of microscopic world and the apparent continuity of the macroscopic
universe.

The moments of junction map a timeless Poincar\'{e} section
representing a state of the involved spaces into another state.
Each Poincar\'{e} section may present some relationship with what
t'Hooft (1999) called Cauchy surfaces of equal time. The moments
of junction represent the interval between two successive states
(each timeless) of a universe. Let E(i) be a Poincar\'{e} section
like S(i) defined above: if it is an identity mapping, $\rm
MJ=Id(S)$, then there is no time interval from S(i) to $\rm
S(i+1)$. In all other cases, the MJ represents two important
parameters: first, it accounts for a differential time interval,
and then for a differential element of the geometry of the
corresponding space. In this sense, it has neither "thickness" nor
duration. There is no "distance" in the Hausdorff sense between
S(i) and $\rm S(i+1)$, just a change in the topological situation.
Since the step from S(i) to $\rm S(i+1)$ is a discrete one, it
follows that: (i) the corresponding space owns discrete that is
likely quantum properties and (ii) these discrete properties are
valid whatever the scales, since they are founded on the set
difference that is not dependent on any scale nor size of
phenomena. It is noteworthy that these properties meet some
requirement for space, time and matter, as suggested by t'Hooft
(1999).

\medskip
\noindent \textbf{8.2. Space-time and motion}

\medskip
The moment of junction formalizes the topological characteristics
of what is called {\bf \textit{motion}}, in a physical universe,
that is what has been considered as needed for the understanding
of physics (Rothwarf, 1998). While an identity mapping denotes an
absence of motion, that is a null interval of time, a nonempty
moment of junction stands for the minimal of any time interval:
this meets a proposition of Sidharth (1999), stating that there
should exist a minimum space-time interval and that "one cannot go
to arbitrarily small space-time intervals or points". In our
sense, there is no such "point": only instants which per se do not
reflect timely features.

The need for morphisms of the topological structures as frame of
reference as combined with the morphisms of objects interestingly
meets with a requirement hypothesized by t'Hooft (1999): "beables"
as commutable operators might be represented in morphisms of
frames and the moments of junction, while "changeables" as non
commutable components might be related to morphisms of objects, up
to the not commutative biological components, which makes the
transition to the next points.

Furthermore, morphisms are compatible with rotations that are to
be expected in eddy-motion which stands at the basis of physical
motion in a nonlinear universe (Wu and Lin, 2002). It should be
noted that discontinuities may occur from changes occurring when a
structure partly escapes the observable space-time through
dimensional plunging into the embedding 4-space and re-emerges in
the perceived 3-space.

\newpage

\noindent \textbf{8.3. Space-time reversibility properties}

\medskip

It is noteworthy that the moment of junction is provided with reversibility,
which accounts for the temporal reversibility of physical phenomena, as
postulated by cosmologists (Smaller and Temple, 2000). In contrast, the
biological arrow of time exhibits some irreversible features, since the
sequence of brain mental images is founded on surjective mappings, which do
not have the same fixed points in the reverse sense (Bounias, 2000b): this
means that even if a biological system is physically reversible, the
correspondence of mental images associated to outside perceptions with the
the self of the perceiver would be changed.

\medskip
\noindent \textbf{8.4. Topological constraints on space-time}

\medskip

Seriu (2000) proposed a very interesting study on metric
properties of a space of spaces. The author reaches the conclusion
that physical constraints suggest the existence of drastic
topological fluctuations at Planck scales. These observations
would result in two correlative hypothesis: first, there would be
sets of spaces with various topologies, and second, there would be
scale-dependent topologies. Seriu acknowledges the risk that such
considerations depart from the very foundations of topology as a
mathematical concept. However, a mathematical space can give raise
to several topologies, which range from coarser to finer forms, in
an order relation (Bourbaki, 1990). In contrast with the smoothing
at large scale recalled by t'Hooft (1999), here a smoothing of
topologies at low scale would be needed. These apparent
contradictions vanish with the properties of the empty hyperset
providing discrete features at all scales, but also own the power
of continuum, that is physical "continuity" inside each
fundamental cell. Note that continuity in the mathematical sense
does not require smoothing. Concerning the problem of
scale-related topological changes, it should be pointed that the
set-distance is a scale-independent measure, and would thus
fulfill a requirement formerly raised in Part 1 of this study
(Bounias and Krasnoholovets, 2003), in the form of an ultrafilter
of the topologies required by Seriu. No contradiction seems to
lurk in these approaches.

The spectra $\Delta^{\infty}$ proposed by Seriu should be
invariant by spatial diffeomorphism: this implies continued
differentiability, a property which is fulfilled by the
convolution structure derived from our model. In addition, while
the Laplacian used by Seriu as a probe for the geometry of the
explored universe accounts for linear properties, the nonlinear
convolution provides a generalization to nonlinear properties.
While scale-dependent topologies may appear contradictory with
fractal properties, our approach instead is consistent with such a
structure

\medskip

\noindent \textbf{8.5. Fractal space-time features}

\medskip

Following the pioneer remarks of Mandelbrot (1989), some fractal
properties have been found for the distribution of galaxies at
rather small scales. However, the existence of a lower cut-off, as
shared by the an spectra of Seriu (2000) seen above, precludes
that the observed autocorrelations reflect a general fractality of
the entire universe. In our model, the lattice that provides space
and matter with their properties has been shown to meet the
properties of a true fractal. This means that some fractal
features should be shared by objects at all scales of our
space-time. The contradiction of uniformity of Hubble law of
isotropy of the background waves with the heterogeneity of a
fractal geometry exhibiting voids and structures is just apparent.
When Mandelbrot drew fractal galaxies on a sheet of paper, the
"void" parts were represented by some lines connecting groups of
points on the sheets. This would pose a problem only if an
absolute void would be postulated between objects in the universe.
In contrast, our approach suggests that a common fractal structure
may hold on the whole of space as an embedding (nonmaterial)
medium. It has been sketched here how this fractal geometry can
account for the formation of matter corpuscles: due to all-scales
self-similarity ratios, the same property should be extended to
large scale formations. This would predict that similar fractal
parameters should be found for both particle scales and cosmic
scales. Lastly, that objects are constructed in and from a fractal
lattice suggests that fractality if provided by the embedding
medium: this answers a question raised by Cherbit (1987) and
supports a former proposition of Feynman (1965) that the quantum
trajectory of a particle is continued but not derivable.

\medskip
\noindent \textbf{8.6. Predictions of fundamental physical
parameters}

\medskip

In the absence of "given" knowledge of what an object could be,
Krasnoholovets and Ivanovsly (1993), Krasnoholovets (1997,
2000a,b, 2001a,b,c) and Krasnoholovets and Byckov (2000) proposed
that a corpuscle could be represented by a local change in the
geometry of a lattice. Independently, Bounias and Bonaly (1995,
1996, 1997) studied why a mathematical space could exist and how
it could provide existence to a physical-like space. It came that
the results support both the hypothesis of existence of a founding
lattice of Krasnoholovets et al., and the prediction of emergence
of the phenomenon of self-conscious perception, which could stand
for a major characteristic of life (Bounias and Bonaly, 1997b,
Bounias, 2000). Since the model makes useless the discrimination
between relativistic and quantum approaches, we modestly expect
that it might be fruitfully thought-provoking to the community,
with emphasis on the fact that many various theories harbour
correct elements that presently are diluted in a complex network
of scattered hypothesis.

The next part of this study (Part 3) will provide a formal
description of the construction and structures of the various
fundamental particles allowed by the model and examine some
implications up to cosmic scales and the origins and behavior of
universes.

\section*{References}

\noindent Blokhintsev, D., 1976. {\it Principles of quantum
mechanics}, Russian, French edition 1981, MIR, Moscow, 683 pp. \\

\noindent Bolivar-Toledo, O., Candela Sola, S., Munoz Blanco,
J.A., 1985. "Nonlinear data transforms in perceptual systems",
{\it Lecture Notes in Computer Sciences}, Vol. 410, pp. 1-9.   \\

\noindent Bonaly, A., 1992. Personal communication to M. Bounias.
\\

\noindent Bonaly, A., Bounias, M., 1995. "The trace of time in
Poincare sections of topological spaces", {\it Physics Essays},
Vol. 8, No. 2, pp. 236-44.   \\

\noindent Bounias, M., 1997. "Definition and some properties of
set-differences, instances and their momentum, in the search for
probationary spaces", {\it Ultra Scientist of Physical Sciences},
Vol. 9, No. 2, pp. 139-45.   \\

\noindent Bounias, M., 2000a. "The theory of something: a theorem
supporting the conditions for existence of a physical universe,
from the empty set to the biological self", in Daniel M. Dubois,
(Ed.) CASYS'99 Int. Math. Conf.,  {\it Int. J. Computing
Anticipatory Systems}, Vol. 5, pp. 11-24.   \\

\noindent Bounias, 2000b. "A theorem proving the irreversibility
of the biological arrow of time, based on fixed points in the
brain as a compact or delta-complete space", {\it Am. Inst. Phys.
Conf. Proc.}, Vol. CP517, pp. 233-43. \\

\noindent Bounias, M., 2001. "Indecidability and Incompleteness In
Formal Axiomatics as Questioned by Anticipatory Processes", in
Daniel M. Dubois (Ed.) CASYS'2000 Int. Math. Conf., {\it Int. J.
Computing Anticipatory Systems} (in press).  \\

\noindent Bounias, M., Bonaly, A., 1996. "On metrics and scaling :
physical coordinates in topological spaces", {\it Indian Journal
of Theoretical Physics}, Vol. 44, No. 4, pp. 303-21.   \\

\noindent Bounias, M., Bonaly, A., 1997a. "The topology of
perceptive functions as a corollary of the theorem of existence in
closed spaces", {\it BioSystems}, Vol. 42, pp. 191-205.  \\

\noindent Bounias, M., Bonaly, A., 1997b. "Some theorems on the
empty set as necessary and sufficient for the primary topological
axioms of physical existence", {\it Physics Essays}, Vol. 10, No. 4,
pp. 633-43.  \\

\noindent Bounias, M., Krasnoholovets, V., 2003. "Scanning the
structure of ill-known spaces: Part 1. Founding principles about
mathematical constitution of space", {\it Kybernetes: The Int. J.
of Systems \& Cybernetics}, Vol. 32, No. 7/8, pp. 945-75
(also physics/0211096).  \\

\noindent Bourbaki, N., 1990a. {\it Theorie des ensembles},
Masson, Paris, Chapters 1-4, p. 352.   \\

\noindent Bourbaki, N., 1990b. {\it Topologie Gendrale}, Masson,
Paris, Chapters 1-4, p. 376.   \\

\noindent Boyer, T., 2000. "The infinitely empty does not exist".
{\it Pour La Science} ({\it Scientific American}, French edition),
Vol. 278, pp. 128-37.    \\

\noindent Bucher, M., Spergel, D., 1999. "L'inflation de 1'
univers", {\it Pour La Science} ({\it Scientific American}, French
edition), Vol. 257, pp. 50-7.   \\

\noindent Cherbit, 1987. "Dimension locale, quantite de mouvement
et trajectoires", in {\it Fractals}, Masson, Paris, 340-52.   \\

\noindent Duff, M., 1998. "The new string theory", {\it Pour La
Science} ({\it Scientific American}, French edition), Vol. 246, p. 68.   \\

\noindent Einstein, A., 1920. {\it Aether and the theory of
relativity}. Leyden University Lecture. French translation:
Gauthier-Villars, Paris, pp. 1-12 (1992). \\

\noindent Feynman, R., Hibbs, A., 1965. {\it Quantum mechanics and
path integrals}, McGraw Hill, NY, p. 176.  \\

\noindent Keilman, Y., 1998. "On the breakdown of the principle of
relativity", {\it Physics Essays}, Vol. 11, No. 2, pp. 325-29.  \\

\noindent Kozyrev, N.A., Nasonov, V.V., 1978. In {\it Astronometry
and celestian mechanics}, Akademia Nauk SSSR, Moscow, Leningrad,
pp. 168-79 (in Russian).  \\

\noindent Krasnoholovets, V., Ivanovsky, D., 1993. "Motion of a
particle and the vacuum", {\it Physics Essays}, Vol. 6, No. 4,
554-63. (Also arXiv.org e-print archive quant-ph/9910023).   \\

\noindent Krasnoholovets, V., 1997. "Motion of a relativistic
particle and the vacuum", {\it Physics Essays}, Vol. 10, No. 3,
pp. 407-16. (Also quant-ph/9903077). \\

\noindent Krasnoholovets, V., Byckov, V., 2000. "Real inertons
against hypothetical gravitons. Experimental proof of the
existence of inertons", {\it Indian Journal of Theoretical
Physics}, Vol. 48, No. 1, pp. 1-23. (Also
quant-ph/0007027).  \\

\noindent Krasnoholovets, V., 2000. "On the nature of spin,
inertia and gravity of a moving canonical particle", {\it Indian
Journal of Theoretical Physics}, Vol. 48, No. 2, pp. 97-132.
(Also quant-ph/0103110).  \\

\noindent Krasnoholovets, V., 2001a. "On the theory of anomalous
photoelectric effect stemming from a substructure of matter
waves", {\it Indian Journal of Theoretical Physics}, Vol. 49, No. 1,
pp. 1-32. (Also quant-ph/9906091).  \\

\noindent Krasnoholovets, V., 2001b. "Space structure and quantum
mechanics", {Space-time \& Substance}, Vol. 1, No. 4, pp. 172-75.
(Also quant-ph/0106106).  \\

\noindent Krasnoholovets, V., 2001c. "On the way to submicroscopic
description of nature",  {\it Indian Journal of Theoretical
Physics}, Vol. 49, No. 2, 81-95. (Also quant-ph/9908042).  \\

\noindent Krasnoholovets, V., 2001d. "Collective dynamics of
hydrogen atoms in the KIO$_3\cdot$HIO$_3$ crystal dictated by a
substructure of the hydrogen atoms' matter waves",
cond-mat/0108417. \\

\noindent Krasnoholovets, V., Strokach, O., Skliarenko,   Akimov,
L. 2001. {\it Inerton astronomy: Facility for measuring the
inerton radiation from stars and planets}, project.  \\

\noindent Lin, Y., Wu, Y., 1998. "Blown-ups and the concept of
whole evolutions in systems science", {\it Problems of Nonlinear
Analysis in Engineering Systems}, Vol. 4, pp. 16-31.  \\

\noindent Maldacena, J., 2000. "Gravity, particle physics and
their unification", {\it eConf C990809}, pp. 840-52.
(Also arXiv.org e-print archive \ hep-th/0002092). \\

\noindent Marinov, S., 1996. "Cosmological aspects of the absolute
space-time theory", {\it Physics Essays}, Vol. 9, No. 3,
pp. 357-67.  \\

\noindent Okun, L.B., 1988. {\it Physics of elementary particles},
Nauka, Moscow (in Russian), p. 92. \\

\noindent Ralston, J., Jain, P., Nodland, B., 1997. "The corscrew
effect", {\it Phys. Rev. Lett.}, Vol. 816, pp. 26-9. \\

\noindent Ralston, J.P., Nodland, B., 1997. "An update on
cosmological anisotropy in electromagnetic propagation", in
Donnelly T.W. (Ed.) {\it Proc. 7th Internat. Conf. on the
Intersection of Particle and Nuclear Physics}, Big Sky, Montana,
Amer. Inst. Phys. CP .(Also astro-ph/978114.) \\

\noindent Sidharth, B.C., 1999. "The fractal universe: from the
Planck scale to the Hubble scale", quant-ph/9907024.  \\

\noindent Seriu, M., 2000. "Space of spaces as a metric space",
{\it Commun. Math. Phys.}, Vol. 209, pp. 393-405.  \\

\noindent Smoller, J., Temple, B., 2000. "Cosmology with a
shock-wave", {\it Commun. Math. Phys.}, Vol. 210, pp. 275-308. \\

\noindent t'Hooft, G., 1999. "Quantum gravity as a dissipative
deterministic system", {\it Class. Quantum Grav.}, Vol. 16,
pp. 3263-279. \\

\noindent Walker, F.L., 1996. "The expanding space paradox: can
the galaxies really recede?", {\it Physics Essays}, Vol. 9, No. 2,
pp. 209-15. \\

\noindent Wu, Y., Lin, Y., (2002). "Beyond nonstructural
quantitative analysis", in {\it Blown ups, spinning currents and
modern science}, World Scientific, New Jersey, London, p. 324. \\

\subsection*{Further reading}

\noindent Bounias, M., Bonaly, A., 1994. "On mathematical links
between physical existence, observability and information: towards
a "theorem of something", {\it Ultra Scientist of Physical
Sciences}, Vol. 6, No. 2, pp. 251-9.   \\

\noindent Hales, T.C., 2000. "Cannonballs and honeycombs", {\it
Notices of the AMS}, Vol. 47, No. 4, pp. 440-9.  \\

\noindent Hannon, R.J., 1998. "An alternative explanation of the
cosmological redshift", {\it Physics Essays}, Vol. 11, No. 4,
pp. 576-8.  \\

\noindent Havard, G., Zinsmeister, M., 2000. "Thermodynamic
formalism and variations of the Hausdorff dimension of quadratic
Julia sets", {\it Commun. Math. Phys.}, Vol. 210, pp. 225-47.  \\

\noindent Joyce, M., Anderson, P.W., Montuori, M., Pietronero, L.,
Sylos Labini, F., 2000. "Fractal cosmology in an open universe",
{\it Europhysics Letters}, Vol. 49, pp. 416-22. \\

\noindent Lavrentiev, M.M., Eganova, I.A., Lutset, M.K., Fominykh,
S.F., 1990. "About distance influence of stars on resistor", {\it
Proc. Acad. Sci. USSR}, Vol. 314, No.
2, pp. 352-5 (in Russian). \\

\end{document}